\documentclass[fleqn,10pt]{wlscirep_}
\usepackage[LGR,T1]{fontenc}
\usepackage{alphabeta}
\usepackage{amsmath}
\usepackage{booktabs}
\usepackage{tabularx}
\usepackage{multirow}
\usepackage{enumitem}
\usepackage{stackengine}
\usepackage{graphicx}
\DeclareMathOperator*{\argmin}{arg\,min}
\newcommand\aggsym{\alpha}
\newcommand\analysisDate{May 24, 2020}

\title{Tracking COVID-19 using online search}

\author[1,*]{Vasileios Lampos}
\author[2,3]{Maimuna S. Majumder}
\author[4]{Elad Yom-Tov}
\author[5]{Michael Edelstein}
\author[1]{Simon Moura}
\author[6]{Yohhei Hamada}
\author[6,7]{Molebogeng X. Rangaka}
\author[8,9]{Rachel A. McKendry}
\author[1,10]{Ingemar J. Cox}
\affil[1]{Department of Computer Science, University College London, London, UK}
\affil[2]{Computational Health Informatics Program, Boston Children's Hospital, Boston, MA, USA}
\affil[3]{Department of Pediatrics, Harvard Medical School, Boston, MA, USA}
\affil[4]{Microsoft Research, Herzeliya, Israel}
\affil[5]{National Infection Service, Public Health England, London, UK}
\affil[6]{Institute for Global Health, University College London, London, UK}
\affil[7]{Division of Epidemiology and Biostatistics, University of Cape Town, Cape Town, South Africa}
\affil[8]{London Centre for Nanotechnology, University College London, London, UK}
\affil[9]{Division of Medicine, University College London, London, UK}
\affil[10]{Centre for Communication and Computing, University of Copenhagen, Copenhagen, Denmark\vspace{0.15in}}
\affil[*]{Corresponding author, v.lampos@ucl.ac.uk
\vspace{0.2in}

\textbf{Competing interests:} The authors declare that there are no competing interests.
\vspace{0.3in}
}


\begin{abstract}
Previous research has demonstrated that various properties of infectious diseases can be inferred from online search behaviour. In this work we use time series of online search query frequencies to gain insights about the prevalence of COVID-19 in multiple countries. We first develop unsupervised modelling techniques based on associated symptom categories identified by the United Kingdom's National Health Service and Public Health England. We then attempt to minimise an expected bias in these signals caused by public interest ---as opposed to infections--- using the proportion of news media coverage devoted to COVID-19 as a proxy indicator. Our analysis indicates that models based on online searches precede the reported confirmed cases and deaths by 16.7 (10.2 -- 23.2) and 22.1 (17.4 -- 26.9) days, respectively. We also investigate transfer learning techniques for mapping supervised models from countries where the spread of disease has progressed extensively to countries that are in earlier phases of their respective epidemic curves. Furthermore, we compare time series of online search activity against confirmed COVID-19 cases or deaths jointly across multiple countries, uncovering interesting querying patterns, including the finding that rarer symptoms are better predictors than common ones. Finally, we show that web searches improve the short-term forecasting accuracy of autoregressive models for COVID-19 deaths. Our work provides evidence that online search data can be used to develop complementary public health surveillance methods to help inform the COVID-19 response in conjunction with more established approaches.
\end{abstract}

\begin{document}

\flushbottom

{        
\large

\noindent This preprint is now published in Nature Digital Medicine (\href{https://www.nature.com/articles/s41746-021-00384-w}{nature.com/articles/s41746-021-00384-w}).

\vspace{0.2in}
        
\noindent\textbf{Cite as:}\\
Lampos, V., Majumder, M.S., Yom-Tov, E. \emph{et al}. Tracking COVID-19 using online search. \emph{Nature Digital Medicine} 4, \textbf{17} (2021). \href{https://doi.org/10.1038/s41746-021-00384-w}{https://doi.org/10.1038/s41746-021-00384-w}

\vspace{0.4in}
        
\noindent \textbf{Please note that the final published version differs from this preprint}.

\noindent This preprint was first published on March 18, 2020 (v1) and was subsequently updated up until July 19, 2020 (v10). v12 was produced to fix an error with the references of the preprint. The paper was accepted for publication on December 24, 2020 after 2 review rounds. It was published online on February 8, 2021.
}

\maketitle

\thispagestyle{empty}

\section*{Introduction}
Over the past several years, numerous scientific studies have shown that user interactions with web applications generate latent health-related signals, reflective of individual as well as community level trends~\cite{polgreen2008using,ginsberg2008flu,eysenbach2009,lampos2010tracking,culotta2010towards,dechoudhury2013,yomtov2016,wagner2018added}. Seasonal influenza and the H1N1 pandemic were used as case studies for the development and evaluation of machine learning models that produce disease rate estimates in a non-traditional way, using online search or social media as their input information~\cite{polgreen2008using,ginsberg2008flu,culotta2010towards,chew2010,lampos2012nowcasting}. After an extensive scientific debate about their usefulness and accuracy~\cite{olson2013,Lazer2014}, and with the manifestation of new findings that improved upon past methodological shortcomings~\cite{lampos2015gft,yang2015accurate,lampos2017www}, these techniques are now becoming part of public health systems, serving as complementary endpoints for monitoring the prevalence of infectious diseases, such as seasonal influenza~\cite{wagner2018added,Reich2019}. Compared to conventional health surveillance systems, online user trails exhibit certain advantages, including low latency, continuous operational capacity, denser spatial coverage, and broader demographic inclusion~\cite{lampos2015assessing,wagner2018added}. Furthermore, during emerging epidemics they may offer community level insights that current monitoring systems are not equipped to obtain given a limited testing capacity~\cite{roser2020}, and the physical distancing measures that discourage or prohibit people from interacting with health services~\cite{lipsitch2020covid}. Notably, the ongoing coronavirus disease (COVID-19) pandemic, caused by a novel coronavirus (SARS-CoV-2), has generated an unprecedented relative volume of online searches (Fig.~S1). This search behaviour could signal the presence of actual infections, but may also be due to general concern that is intensified by news media coverage, reported figures of disease incidence and mortality across the world, and imposed physical distancing measures~\cite{baumgartner2011,Qiu2020}.

Previous research, which has used online search to predominantly model influenza-like illness (ILI) rates, has focused on supervised learning solutions, where ``ground truth'' information, in the form of historical syndromic surveillance reports, can be used to train machine learning models~\cite{ginsberg2008flu,culotta2010towards,lampos2015gft,santillana2015,yang2015accurate}. These models learn a function that maps time series of online user-generated data, e.g. the frequency of web searches over time, to a noisy representation of a disease rate time series, indirectly minimising the impact of online activity that is not caused by infection. Typically, this training data spans multiple years and flu seasons~\cite{lampos2015gft,yang2015accurate,lampos2017www}. However, for most locations, if not all, no sufficient data ---in terms of validity, representativeness, and time span--- currently exists to apply supervised learning approaches to COVID-19. Therefore, unsupervised solutions that attempt to minimise the effect of concern should be sought, and fully supervised solutions should be used and interpreted with caution.

In this paper, we present models for COVID-19 using online search data based in both unsupervised and supervised settings. We first develop an unsupervised model for COVID-19 by carefully choosing search queries that refer to related symptoms as identified by a survey from the National Health Service (NHS) and Public Health England (PHE) in the United Kingdom (UK)~\cite{Boddington2020}. Symptom categories are weighted based on their reported ratio of occurrence in cases of COVID-19. The output from these new models provides useful insights, including early warnings for potential disease spread, and showcases the effect of physical distancing measures. Since online searches can also be driven by concern rather than by infection, we attempt to minimise this effect by incorporating a news media coverage time series. The influence of news media on the online search signal is quantified through an autoregressive task that resembles a Granger causality test between these two time series~\cite{granger1969}.

Recent research has demonstrated that a model estimating ILI rates from web searches, trained originally for a location where syndromic surveillance data is available, can be adapted and deployed to another location that cannot access historical ground truth data~\cite{zou2019www}. The accuracy of the target location model depends on identifying the correct search queries and their corresponding weights via a transfer learning methodology. By adapting this method, we map supervised COVID-19 models from a source to a target country, in an effort to transfer noisy knowledge from areas that are ahead in their epidemic curves to areas that are at earlier stages. This supervised approach transfers the clinical reporting biases of the source location, but given the statistical supervision, it is not affected by concern to the same degree as the unsupervised models. Our analysis reaffirms the insights of the unsupervised approach, showcasing how early warnings could have been obtained from locations that had already experienced the impact of COVID-19. 

Taking noisy supervision a step further, we conduct a correlation and regression analysis to uncover potentially useful online search queries that refer to underlying behavioural or symptomatic patterns in relation to confirmed COVID-19 cases. We show that generic COVID-19-related terms and less common symptoms are the most capable predictors. Finally, we use the task of forecasting confirmed COVID-19 deaths to illustrate that web search data, when added to an autoregressive forecasting model, significantly reduces prediction error.

We present results for a multilingual and multicultural selection of countries -- namely, the United States of America (US), UK (including a comparative analysis for England), Australia, Canada, France, Italy, Greece, and South Africa.

\section*{Results}
\begin{figure}[!t]
    \centering
    \begin{tabular}{cc}
    \multirow{4}{*}[-1.1em]{\rotatebox[origin=c]{90}{\sffamily\small Normalised online search score for COVID-19}} & \def\stackalignment{l}
    \topinset{}{\includegraphics[height=1.2in, clip=true, trim=0.23in 0.35in 0 0]{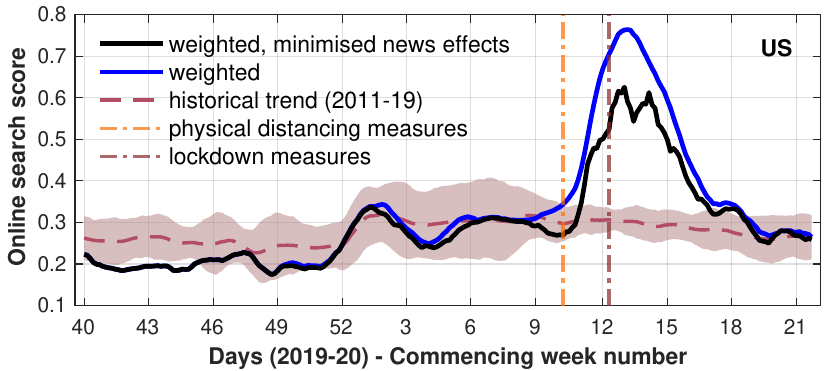}}{0in}{0in}
    \def\stackalignment{l}
    \topinset{}{\hspace{0.04in}\includegraphics[height=1.2in, clip=true, trim=0.45in 0.35in 0 0]{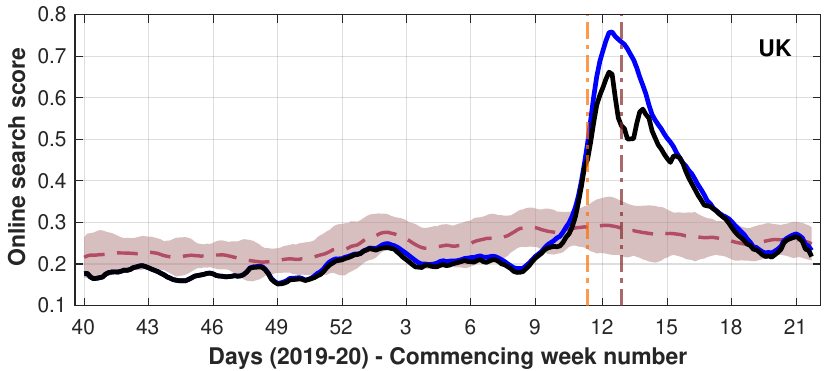}}{0in}{0in} \\
    
    & \def\stackalignment{l}
    \topinset{}{\includegraphics[height=1.2in, clip=true, trim=0.23in 0.35in 0 0]{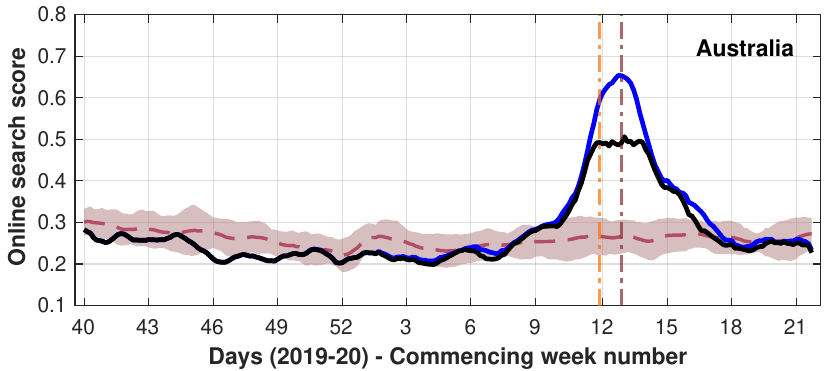}}{0in}{0in}
    \def\stackalignment{l}
    \topinset{}{\hspace{0.04in}\includegraphics[height=1.2in, clip=true, trim=0.45in 0.35in 0 0]{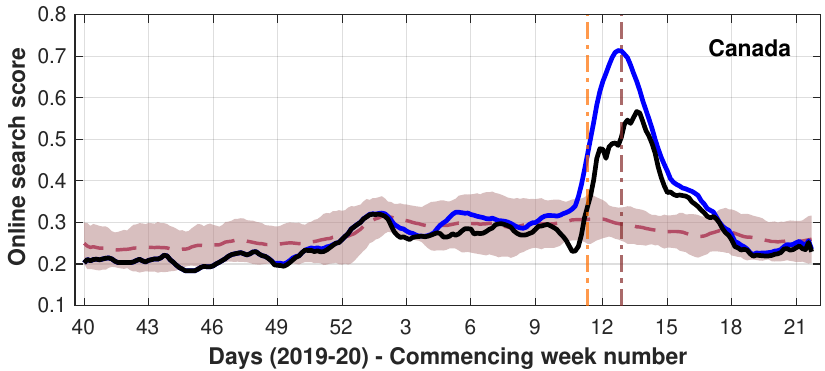}}{0in}{0in} \\
    
    & \def\stackalignment{l}
    \topinset{}{\includegraphics[height=1.2in, clip=true, trim=0.23in 0.35in 0 0]{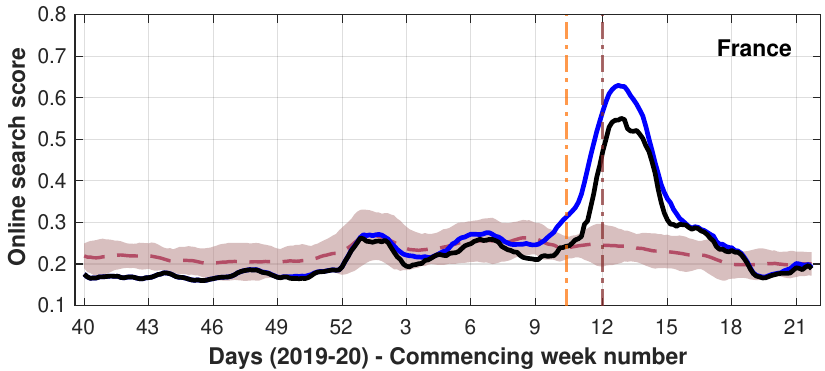}}{0in}{0in}
    \def\stackalignment{l}
    \topinset{}{\hspace{0.04in}\includegraphics[height=1.2in, clip=true, trim=0.45in 0.35in 0 0]{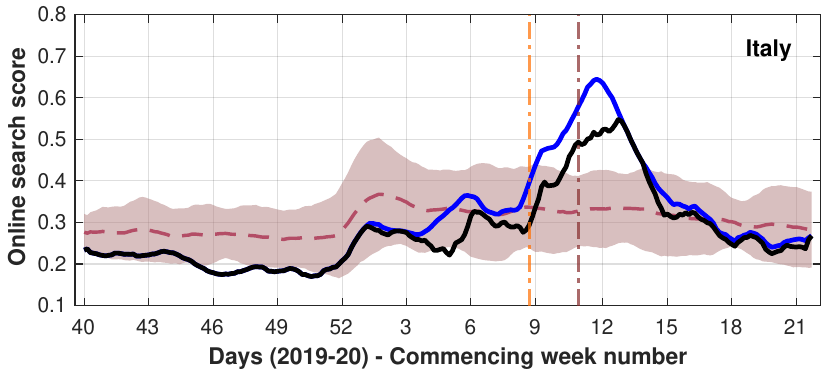}}{0in}{0in} \\
    
    & \def\stackalignment{l}
    \topinset{}{\includegraphics[height=1.285in, clip=true, trim=0.23in 0.20in 0 0]{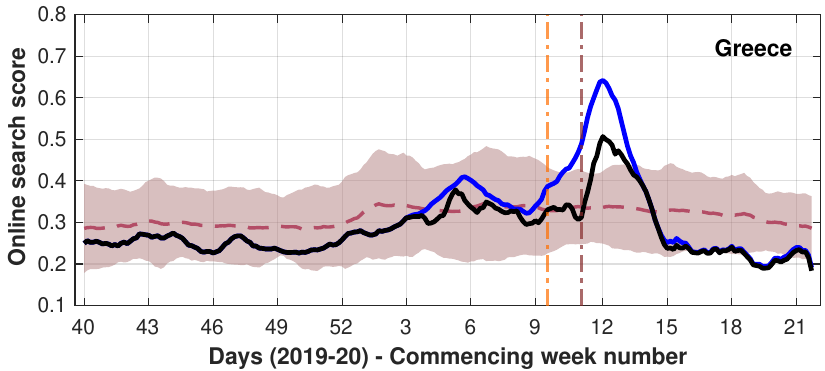}}{0in}{0in}
    \def\stackalignment{l}
    \topinset{}{\hspace{0.04in}\includegraphics[height=1.285in, clip=true, trim=0.45in 0.20in 0 0]{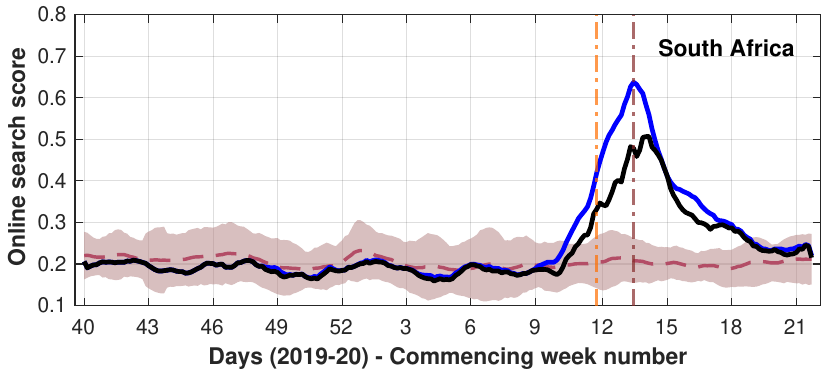}}{0in}{0in}\\
    
    \multicolumn{2}{c}{\sffamily\small\hspace{0.3in} Days (2019-20) -- Commencing week number}
    
    \end{tabular}
    
    \caption{Online search scores for COVID-19-related symptoms as identified by the FF100 survey, in addition to queries with coronavirus-related terms, for $8$ countries from September 30, 2019 to \analysisDate~(all inclusive). Query frequencies are weighted by symptom occurrence probability (blue line) and have news media effects minimised (black line). These scores are compared to an average $8$-year trend of the weighted model (dashed line) and its corresponding $95\%$ confidence intervals (shaded area). Application dates for physical distancing or lockdown measures are indicated with dash-dotted vertical lines; for countries that deployed different regional approaches, the first application of such measures is depicted. All time series are smoothed using a $7$-point moving average, centred around each day.}
    \label{fig:results_ext}
\end{figure}

The first part of the analysis presents unsupervised models of COVID-19 based on weighted search query frequencies. Queries and their weightings are determined using the first few hundred (FF100) survey on COVID-19 conducted by the NHS/PHE in the UK~\cite{Boddington2020}. FF100 identified $19$ symptoms associated with confirmed COVID-19 cases and their probability of occurrence. In addition to these symptoms, we also include queries that mention COVID-19-related keywords (e.g. ``covid-19'' or ``coronavirus'') as a separate category. Unsupervised models for COVID-19 in 8 countries are depicted in Fig.~\ref{fig:results_ext}. We observe exponentially increasing rates that exceed the estimated seasonal average (previous 8 years) during their peak period in all investigated countries, as well as a steep drop of the score after the application of physical distancing or lockdown measures in most countries which is in concordance with clinical surveillance reports~\cite{lau2020}. In the time series where we have attempted to minimise the effect of news by maintaining a proportion of the signal (Eqs.~\ref{eq:news_db_ar},~\ref{eq:news_db_ar2},~and~\ref{eq:news_db_ar_final}), we observe more conservative estimates in all locations, including altered trends during the peak period. Compared to the original scores (no minimisation of news media effects), there is an average reduction of the signals by $16.4\%$ ($14.2\%-18.7\%$) in a period of $14$ days prior and after their peak moments; their corresponding average linear correlation during this period is equal to $.822$ ($.739-.905$). Outside of the peak periods, the reduction is a moderate $3.3\%$ ($2.7\%-4.0\%$), indicating that there is an association between the extent of media impact and the estimated disease prevalence. For the US, Australia, Canada, and Italy, the signal is visibly flattened during the time period around their respective peaks. Countries that took measures earlier in the epidemic curve (e.g. Greece) demonstrate a more pronounced pattern of decrease, with scores that go below the expected seasonal average. We also note that for Australia and the UK, search scores were already in decline after the application of physical distancing measures but before lockdowns. Outcomes based solely on the FF100 symptoms (Fig.~S2) or without using any weighting scheme (Fig.~S3) are available in the Supplementary Information (SI).

\begin{figure}[t]
    \centering
    \begin{tabular}{cc}
    \multirow{4}{*}[-2.2em]{\rotatebox[origin=c]{90}{\sffamily\small Standardised time series trend (z-score)}} & \def\stackalignment{l}
    \topinset{}{\includegraphics[height=1.2in, clip=true, trim=0.23in 0.35in 0 0]{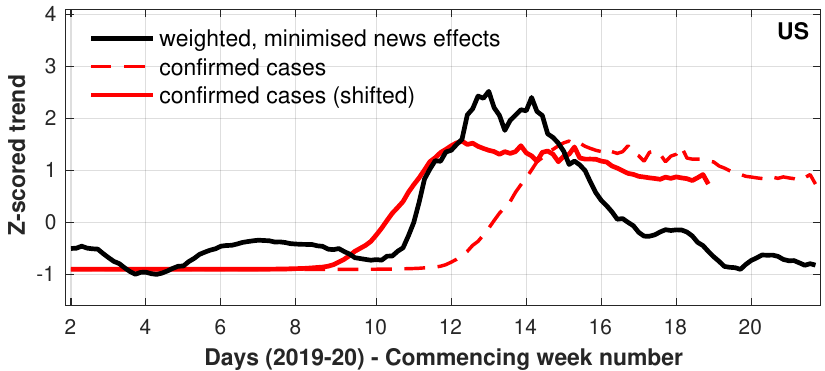}}{0in}{0in}
    \def\stackalignment{l}
    \topinset{}{\hspace{0.04in}\includegraphics[height=1.2in, clip=true, trim=0.38in 0.35in 0 0]{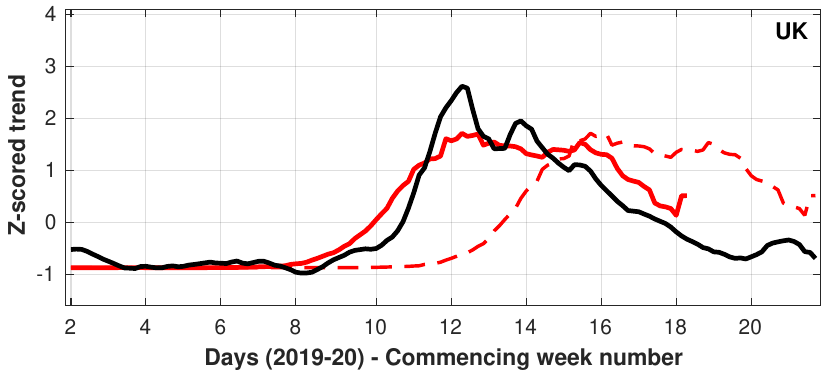}}{0in}{0in} \\
    
    & \def\stackalignment{l}
    \topinset{}{\includegraphics[height=1.2in, clip=true, trim=0.23in 0.35in 0 0]{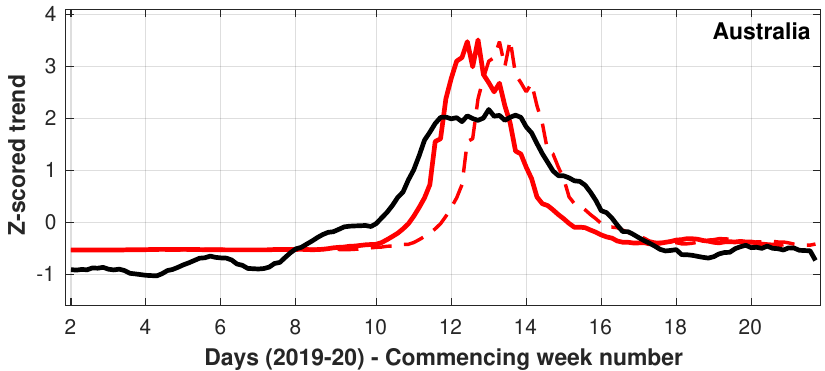}}{0in}{0in}
    \def\stackalignment{l}
    \topinset{}{\hspace{0.04in}\includegraphics[height=1.2in, clip=true, trim=0.38in 0.35in 0 0]{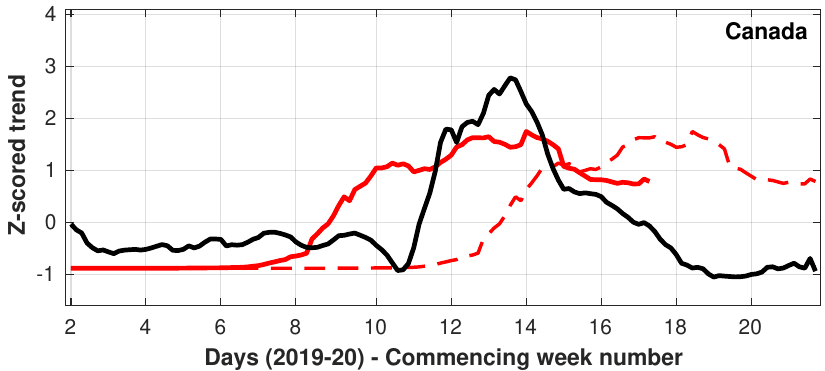}}{0in}{0in} \\
    
    & \def\stackalignment{l}
    \topinset{}{\includegraphics[height=1.2in, clip=true, trim=0.23in 0.35in 0 0]{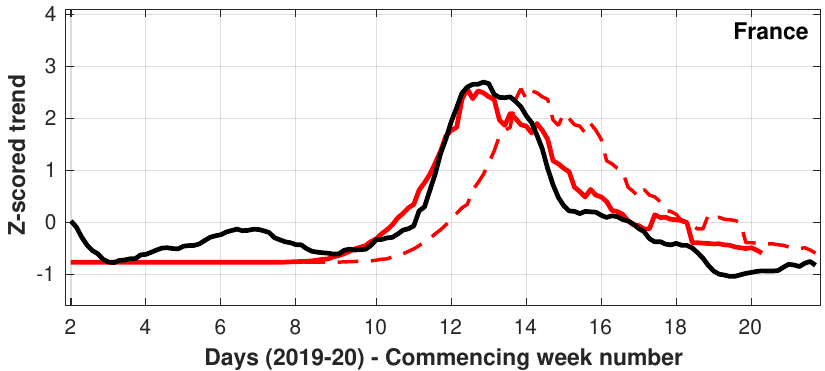}}{0in}{0in}
    \def\stackalignment{l}
    \topinset{}{\hspace{0.04in}\includegraphics[height=1.2in, clip=true, trim=0.38in 0.35in 0 0]{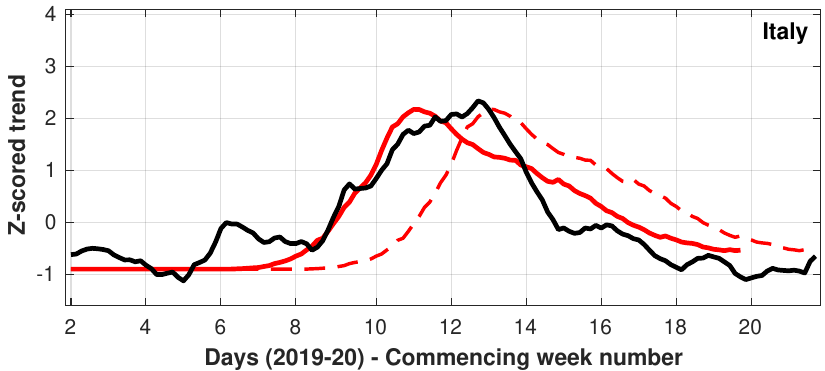}}{0in}{0in} \\
    
    & \def\stackalignment{l}
    \topinset{}{\includegraphics[height=1.285in, clip=true, trim=0.23in 0.20in 0 0]{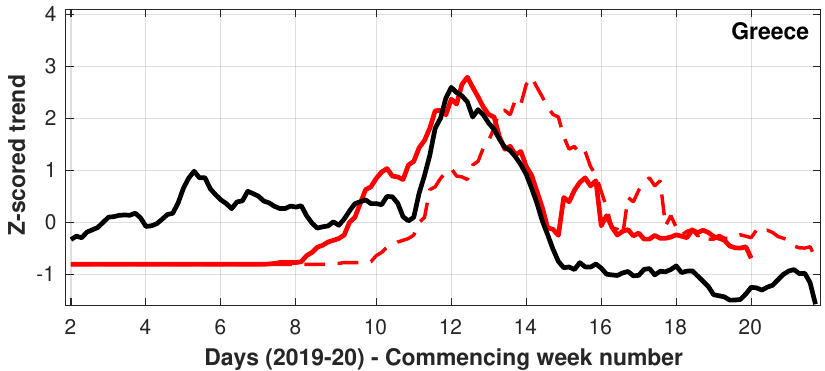}}{0in}{0in}
    \def\stackalignment{l}
    \topinset{}{\hspace{0.04in}\includegraphics[height=1.285in, clip=true, trim=0.38in 0.20in 0 0]{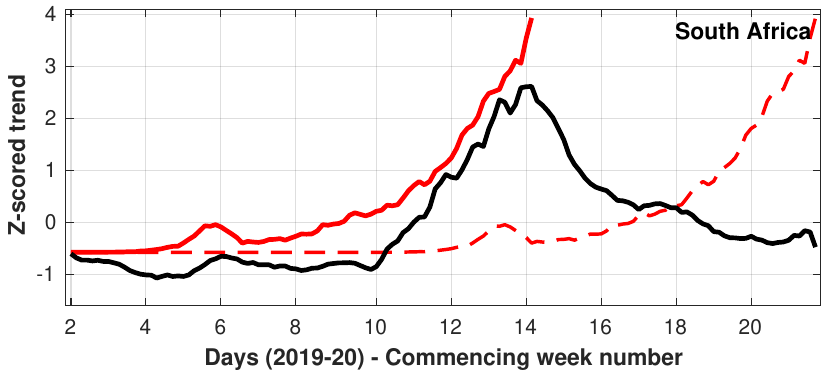}}{0in}{0in}\\
    
    \multicolumn{2}{c}{\sffamily\small\hspace{0.34in} Days (2019-20) -- Commencing week number}
    
    \end{tabular}
    
    \caption{Comparison between online search scores with minimised news media effects (black line) and confirmed cases (dashed red line), as well as confirmed cases shifted back (red line) such that their correlation with the online search scores is maximised. The confirmed cases time series are shifted back by a different number of days for each country: 20 days (US), 24 days (UK), 6 days (Australia), 31 days (Canada), 10 days (France), 14 days (Italy), 12 days (Greece), and 53 days (South Africa). The depicted time interval is from September 30, 2019 to \analysisDate~(all inclusive). All time series are smoothed using a $7$-point moving average, centred around each day.}
    \label{fig:uns_vs_cc}
\end{figure}

A comparison of the search scores with minimised media effects to the time series of confirmed cases is depicted in Fig.~\ref{fig:uns_vs_cc}. If we exclude South Africa, as it displays an outlying behaviour, perhaps due to a limited testing capacity~\cite{hopman2020,gilbert2020} or demographically-skewed internet access patterns~\cite{mahler2019internet}, the correlation between these times series is maximised, reaching an average value of $.826$ ($.735-.917$) when clinical data is brought forward by $16.7$ ($10.2-23.2$) days. This provides an indication of how much sooner the proposed unsupervised models could have signalled an early warning about these epidemics at a national level. Replacing confirmed cases with deaths caused by COVID-19 (Fig.~S4) increases this period to $22.1$ ($17.4-26.9$) days with a slightly greater maximised correlation ($r = .846$; $.702-.990$).

\begin{figure}[t]
    \centering
    \setlength{\tabcolsep}{0.1em}
    \begin{tabular}{ccc}
    
    \def\stackalignment{l}
    \topinset{}{\includegraphics[height=1in, clip=true, trim=0.25in 0.5in 0 0]{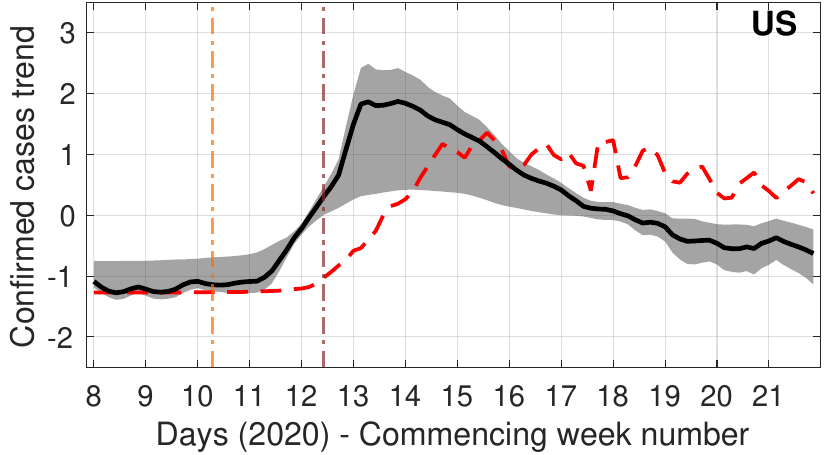}}{0in}{0in} &
    \def\stackalignment{l}
    \topinset{}{\includegraphics[height=1in, clip=true, trim=0.52in 0.5in 0 0]{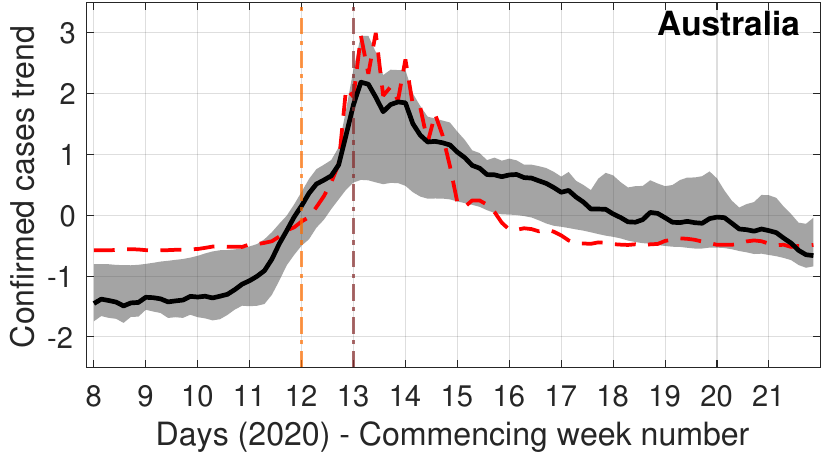}}{0in}{0in} &
    \def\stackalignment{l}
    \topinset{}{\includegraphics[height=1in, clip=true, trim=0.52in 0.5in 0 0]{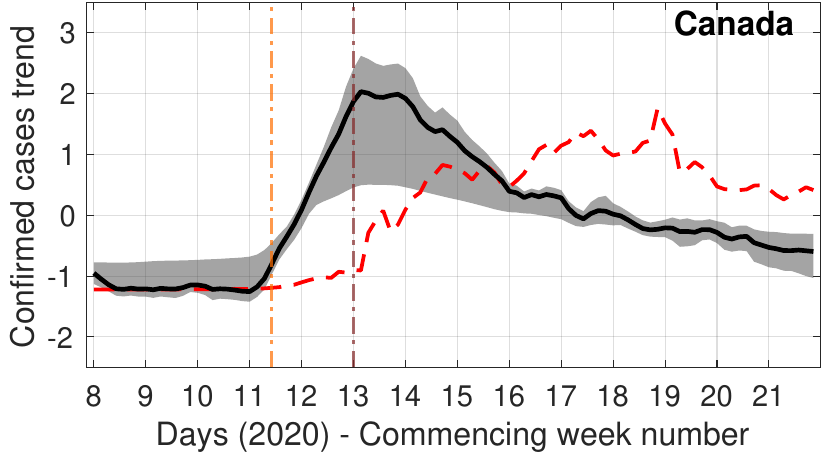}}{0in}{0in} \\
    
    \def\stackalignment{l}
    \topinset{}{\includegraphics[height=1.095in, clip=true, trim=0.25in 0.25in 0 0]{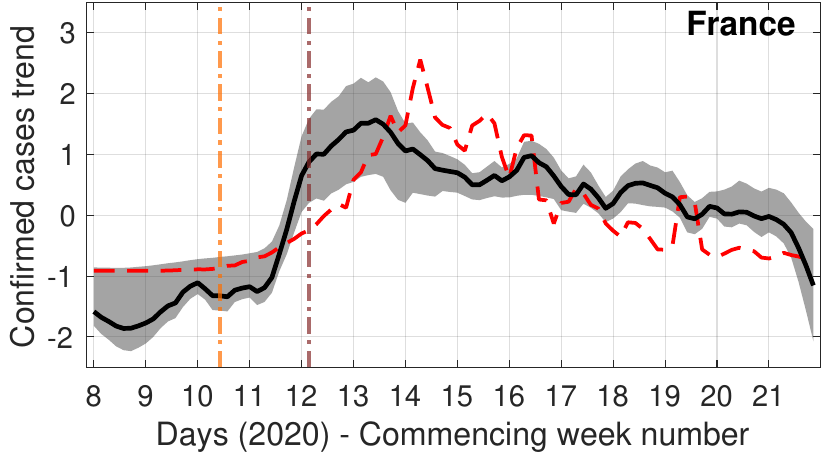}}{0in}{0in} &
    \def\stackalignment{l}
    \topinset{}{\includegraphics[height=1.095in, clip=true, trim=0.52in 0.25in 0 0]{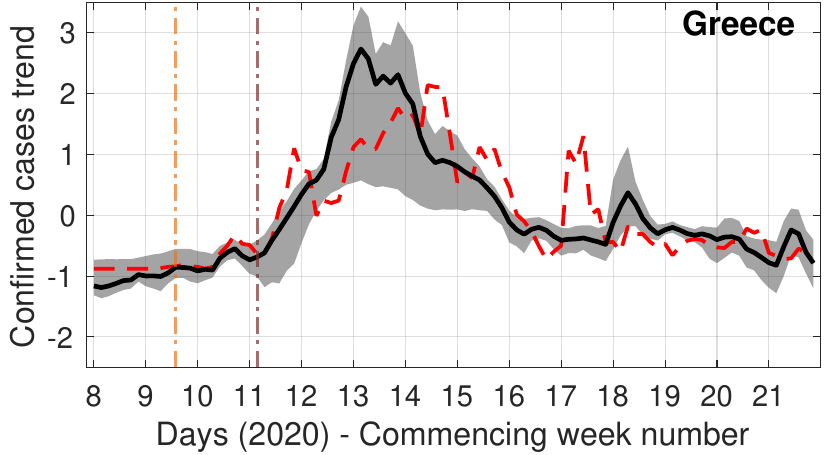}}{0in}{0in} &
    \def\stackalignment{l}
    \topinset{}{\includegraphics[height=1.095in, clip=true, trim=0.52in 0.25in 0 0]{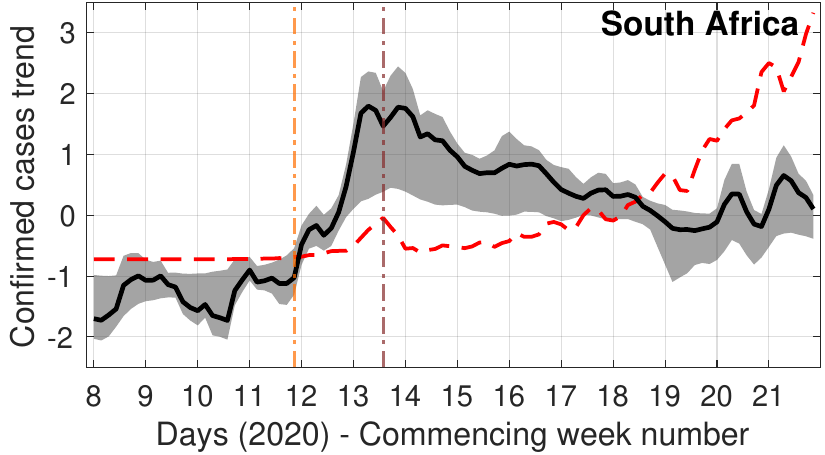}}{0in}{0in} \\
    
    \multicolumn{3}{c}{\def\stackalignment{l}
    \topinset{}{\includegraphics[height=2.1in, clip=true, trim=0 0 0 0]{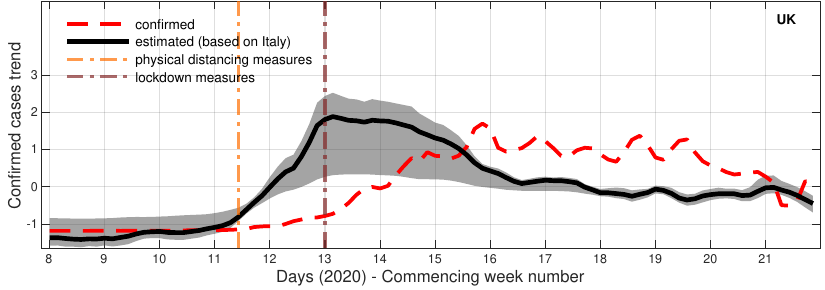}}{0in}{0in}}
    
    \end{tabular}
    
    \caption{Transfer learning models for $7$ countries using Italy as the source country. The figures show an estimated trend for confirmed COVID-19 cases compared to the reported one. The trend is derived by standardising the transferred estimates (raw values are reflective of the demographics and clinical reporting approach of the source country). The solid line represents the mean estimate from an ensemble of models. The shaded area shows $95\%$ confidence intervals based on all model estimates. Application dates for physical distancing or lockdown measures are indicated with dash-dotted vertical lines; for countries that deployed different regional approaches, the first application of such measures is depicted. Time series are standardised and smoothed using a 3-point moving average, centred around each day.}
    \label{fig:results_tl}
\end{figure}

Models of confirmed COVID-19 cases are transferred from Italy (source) to all other countries (targets) in our analysis using a transfer learning methodology. In contrast to the unsupervised models, here we attempt to leverage information from a country that is ahead in terms of epidemic progression~\cite{remuzzi2020}. As a result, the obtained estimates are reflective of the clinical reporting systems in the source country, but not as influenced by user concern at the target countries given that they are derived from a supervised learning function. Our approach maps search queries about specific symptom categories, as identified by the FF100 survey, from the language of the source country (Italian) to the languages of the other countries using a temporal correlation metric (see Methods). We train $100{,}000$ elastic net models for the source location, exploring the entire $\ell_1$-norm regularisation path, and transfer the subset of models ($84{,}557$ on average) that assign a nonzero weight to a minimum of $3$ and up to a maximum of $49$ queries (out of the $54$ queries identified for Italy) on a daily basis from February 17 to May 24, 2020 (both inclusive). The transferred estimates for the last day in the considered time span and their $95\%$ confidence intervals are depicted in Fig.~\ref{fig:results_tl}. Intermediate outcomes showing estimates from models that are trained and transferred on a daily basis are available in Fig.~S5. With the exceptions of Australia and South Africa, the peak of the transferred signal appears approximately $2$-$3$ weeks earlier than the confirmed cases. The decrease after the peak is also more steep in the transferred models and always occurs after the commencement of physical distancing or lockdown measures. The average correlation between the transferred and the unsupervised (with minimised news effects) estimates is equal to $.638$ ($.545-.731$), which implies that there exist differences between the two approaches. In fact, their correlation is maximised by bringing the transferred time series $6$ ($3.63-8.37$) days forward, indicating that the unsupervised signal provides an earlier alert. Interestingly, during the temporal alignment of source and target search query frequencies, which is an intermediate step of the transfer learning technique, correlation increases when the target data is shifted forward. This partially confirms that Italy was indeed ahead by a few days in terms of either epidemic progression, user search behaviour, or both, and justifies our choice to use it as the source country. In particular, when we focus on the period from March 16 to May 24, 2020 (both inclusive) ---dates that signify the beginning and end of high levels of transmission for Italy--- and analyse all transferred models, the average shift in days that maximises these correlations per target country is: $13.76$ ($12.97-14.55$) for the US, $12.67$ ($11.99-13.35$) for the UK, $5.24$ ($4.30-6.18$) for Australia, $8.06$ ($7.14-8.98$) for Canada, $9.84$ ($8.62-11.06$) for France, $1.44$ ($0.45-2.43$) for Greece, and $10.99$ ($10.10-11.88$) for South Africa.

\begin{figure}[t]
    \centering
    \def\stackalignment{l}
    \topinset{\sffamily\textbf{A}}{\includegraphics[height=3.4in, clip=true, trim=0 0 0 0]{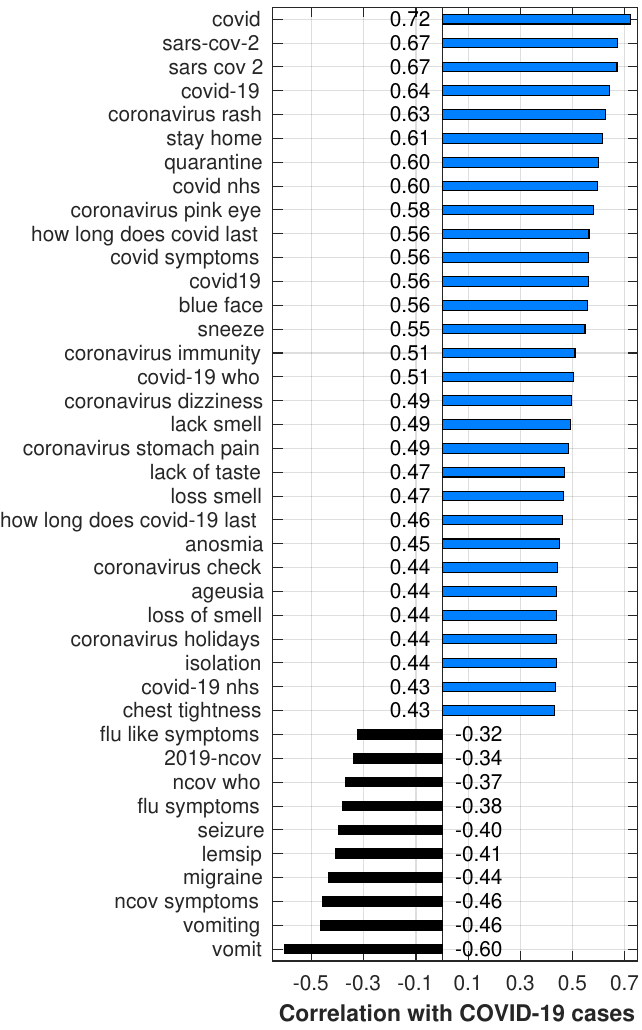}}{0.01in}{0.1in}
    \hspace{0.1in}
    \def\stackalignment{l}
    \topinset{\sffamily\textbf{B}}{\includegraphics[height=3.4in, clip=true, trim=0 0 0 0]{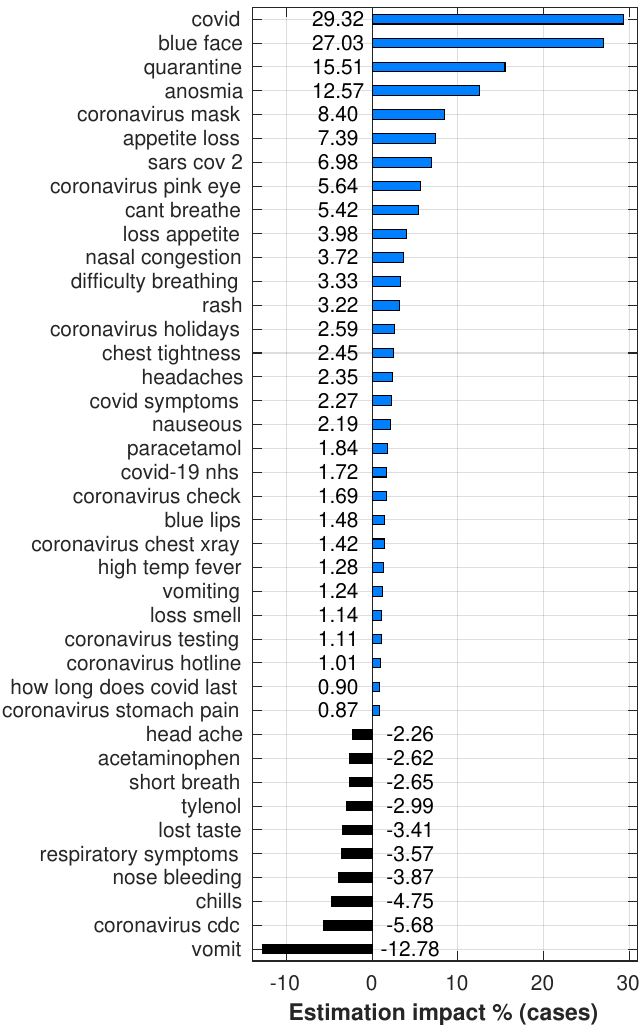}}{0.01in}{0.1in}
    \hspace{0.1in}
    \def\stackalignment{l}
    \topinset{\sffamily\textbf{C}}{\includegraphics[height=3.4in, clip=true, trim=0 0 0 0]{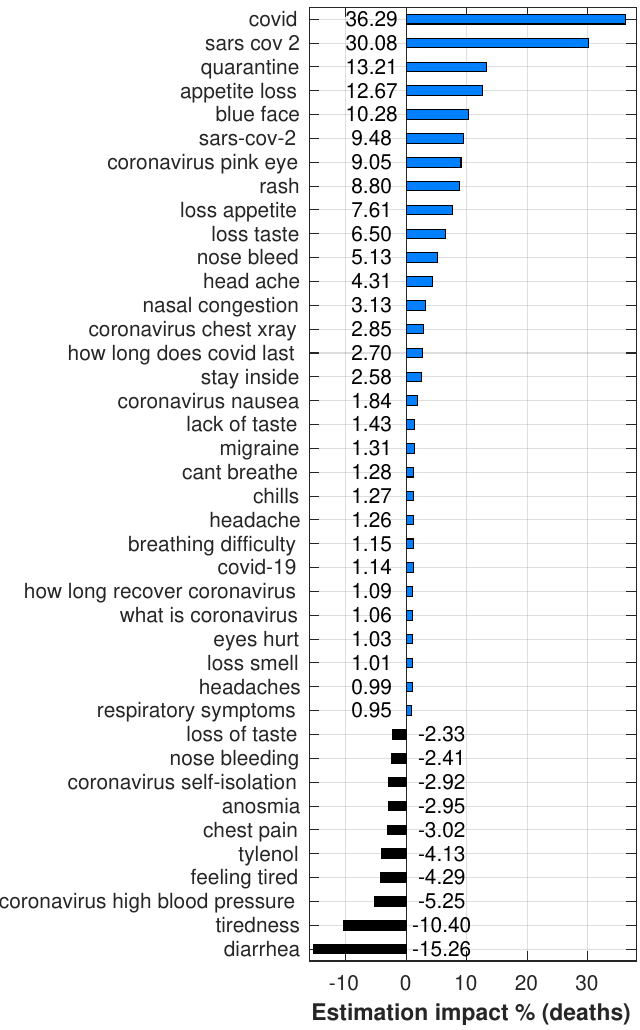}}{0.01in}{0.1in}
    
    \caption{Correlation and regression analysis of search query frequencies against confirmed COVID-19 cases or deaths in four English speaking countries (US, UK, Australia, and Canada). (\textbf{A}) Top-30 positively and top-10 negatively correlated search queries with COVID-19 confirmed cases; (\textbf{B}) Top-30 positively and top-10 negatively impactful queries in estimating COVID-19 confirmed cases; (\textbf{C}) Top-30 positively and top-10 negatively impactful queries in estimating deaths caused by COVID-19.}
    \label{fig:feature_analysis}
\end{figure}

Aiming to uncover symptom-related patterns or associated behaviours, we examine the statistical relationship between web search frequencies and confirmed COVID-19 cases or deaths by performing a correlation and regression analysis. Outcomes could also help inform the choice of search terms in follow-up models for COVID-19. To reduce the representation bias of clinical endpoints, we combine data from multiple countries to the extent possible. For a more comprehensive experiment, we aggregate data from $4$ countries where English is the main spoken language, namely the US, UK, Australia, and Canada. We first estimate the linear correlation between query frequencies and clinical data at all locations (in an aggregate fashion), from December 31, 2019 up to and including \analysisDate. Results illustrating the top-correlated and anti-correlated search queries are depicted in Fig.~\ref{fig:feature_analysis}(A); correlations with deaths, which are lower given the additional temporal difference, are depicted in Fig.~S6(A). Queries about the disease (``covid''; $r = .72$) or the virus (``sars cov 2''; $r = .67$) are the top-correlated. Various related symptoms demonstrate strong correlations as well, although the amount of correlation is not necessarily reflective of the symptom's occurrence probability; examples include rash ($r = .63$), pink eye ($r = .58$), blue face ($r = .56$), sneezing ($r = .55$), and loss of the sense of smell ($r = .49$). Associated behaviours, e.g. staying at home ($r = .61$), or measures, e.g. quarantine ($r = .60$), are also strongly correlated. On the other hand, symptoms such as vomiting ($r = -.60$) and migraine ($r = -.44$) show a significant anti-correlation. When we shift confirmed cases back by $19$ days and deaths by $25$ days, the average correlation across all considered web searches is maximised; results are depicted in Figs.~S6(B)~and~S6(C). The membership and ranking is similar to the previously shown results, although there are marginal increases in the numbers of search queries that seek information about the disease (characteristics, testing, potential treatments). For the regression analysis, we focus on the last $84$ days of the considered time span (March 2 to \analysisDate), training models and assessing estimates for each day of that period. We consider elastic net models with up to a $50\%$ feature density (nonzero weights for half of the considered search queries at most) to minimise the chance of overfitting. The outcomes of this analysis are depicted in Figs.~\ref{fig:feature_analysis}(B)~and~\ref{fig:feature_analysis}(C); for completeness, results where the entire regularisation path is explored ($1\%-100\%$ feature density) are shown in Figs.~S6(D)~and~S6(E). We observe that the most impactful feature, in terms of its normalised contribution to estimates of confirmed cases or deaths, are search queries that include the term ``covid'' ($29.32\%$ and $36.29\%$, respectively), which is intuitive given the magnitude of this pandemic. Blue face ($27.05\%$), loss of the sense of smell ($12.57\%$), appetite loss ($7.39\%$), pink eye ($5.64\%$), and shortness of breath ($5.42\%$) are the top-5 most impactful symptoms with regards to estimating confirmed cases. For mortality estimates, two new symptoms are introduced in the top-5: rash ($8.80\%$) and loss of the sense of taste ($6.5\%$). In addition, recommended behaviours, such as isolation and staying at home, and queries about masks, related diagnostics, or holidays are also impactful. Similarly to the correlation analysis, the search queries related to vomiting have a strong negative impact ($-12.78\%$) in estimating confirmed cases; this is replaced by diarrhoea when estimating deaths ($-15.26\%$). Queries about the common symptoms of cough, fatigue, and fever are not among the most correlated or impactful, suggesting that web searches about rarer symptoms may be more informative.

\begin{table}[!b]
\footnotesize
\renewcommand*{\arraystretch}{1.2}
\setlength{\belowrulesep}{0pt}
\setlength{\aboverulesep}{0pt}
\setlength{\tabcolsep}{5pt}
\centering
\begin{tabular}{l rrr rrr}
\toprule
\multirow{2}{*}{\textbf{Country}} & \multicolumn{3}{c}{\bf 7 days ahead} & \multicolumn{3}{c}{\bf 14 days ahead} \\
&\multicolumn{1}{c}{\bf AR-F} &\multicolumn{1}{c}{\bf SAR-F} &\multicolumn{1}{c}{\bf PER-F} &\multicolumn{1}{c}{\bf AR-F} &\multicolumn{1}{c}{\bf SAR-F} &\multicolumn{1}{c}{\bf PER-F} \\
\midrule
\bf UK & 206.08 (183.25) & 128.37 (115.20) & 137.89 (82.19) & 370.25 (242.63) & 174.77 (181.53) & 227.46 (129.86) \\
\bf US & 774.49 (544.70) & 545.97 (447.31) & 545.06 (667.84) & 1049.07 (697.66) & 591.55 (454.22) & 613.66 (453.73) \\
\bf Australia & 1.48 (0.87) & 0.92 (0.69) & 0.97 (1.01) & 1.14 (0.68) & 0.94 (0.63) & 1.66 (1.35) \\
\bf Canada & 74.61 (46.70) & 55.00 (34.39) & 33.89 (24.47) & 108.48 (50.75) & 82.36 (43.96) & 52.86 (35.79) \\
\bf Greece & 1.76 (1.09) & 1.43 (0.96) & 1.97 (1.64) & 2.14 (1.23) & 1.40 (1.07) & 1.86 (1.78) \\
\bf Italy & 207.21 (97.77) & 85.77 (54.34) & 85.69 (66.30) & 308.99 (134.86) & 162.03 (95.31) & 157.86 (86.09) \\
\bf France & 354.49 (119.69) & 193.81 (120.62) & 151.54 (156.93) & 384.38 (177.43) & 245.85 (147.42) & 282.03 (243.91) \\
\bf South Africa & 6.92 (5.85) & 7.33 (6.53) & 6.14 (5.58) & 9.14 (7.45) & 8.96 (7.25) & 7.68 (6.25) \\
\midrule
\bf Norm. mean & 0.294 (0.192) & 0.198 (0.161) & 0.187 (0.187) & 0.382 (0.241) & 0.253 (0.198) & 0.266 (0.215) \\
\bottomrule
\end{tabular}
\caption{\label{tab:forecastingAC} Average mean absolute error and standard deviation (in parentheses) of forecasting models ($7$ and $14$ days ahead) for daily deaths caused by COVID-19 in 8 countries. The last row contains min-max normalised averages across countries, methods, and forecasting tasks to account for the different ranges in different countries. \textbf{AR-F}: autoregressive forecasting using past deaths; \textbf{SAR-F}: combined online search and autoregressive forecasting; \textbf{PER-F}: persistence model.}
\end{table}

As a final step in our analysis, we use fully supervised forecasting models for COVID-19 deaths to assess whether the inclusion of web searches can improve the accuracy of autoregressive (AR) models. We also provide the same analysis for confirmed cases in the SI (Table~S1, and Fig.~S8), but given the irregularities in the way laboratory diagnostics were conducted, the time series of deaths is more consistent, and hence more appropriate to use for this challenging supervised learning task. We first assess an AR model that uses only deaths data from the past $L = 6$ days (AR-F) to conduct forecasts $7$ and $14$ days ahead. We then expand on this by incorporating online search data as well (SAR-F). Both models are based on Gaussian Processes (GPs) as detailed in Methods. We also use a basic persistence model (PER-F) as a modest baseline. Our testing period starts from April 20, 2020, but we commence testing only when a cumulative number of $10$ deaths is recorded in a country (this reduces the amount of test points for South Africa only). Models are retrained at every time step, and their accuracy is assessed using the mean absolute error (MAE) between forecasts and actual figures. Table~\ref{tab:forecastingAC} enumerates these results, including a normalised average MAE across locations, models, and forecasting tasks to allow for a fairer joint interpretation. We note that SAR-F performs considerably better than AR-F, decreasing MAE by $32.65\%$ and $33.77\%$ in the $7$ and $14$ days ahead forecasting tasks, respectively. It also improves upon the PER-F baseline in the more challenging task, which is a positive outcome given the small amount of available training data. The corresponding $14$ days ahead forecasts are depicted in Fig.~S7. As expected from the empirical evaluation, SAR-F estimates are visibly a better fit to the ground truth than the ones produced by AR-F, capturing the quantity and the overall trend more convincingly. These outcomes provide further evidence for the usefulness of incorporating online search information in disease models for COVID-19.

\section*{Discussion}
We have presented a series of experiments that demonstrate the practical utility of online search in modelling the incidence of COVID-19. Comparing our outcomes to clinical endpoints, we argue that signals from web search data could have served as preliminary early indicators for COVID-19 prevalence at the national level. Our results also highlight the immediate impact that physical distancing or lockdown measures had in reducing disease rates. Furthermore, a qualitative analysis shows that rarer symptoms or generic queries about COVID-19 correlate better with and are more predictive of clinically reported metrics.

Assessing the correctness of our analysis is difficult because there exists no definitive ground truth representative of community level disease rates to compare against. However, intrinsic properties in our findings as well as quantitative comparisons with different clinical endpoints provide at least partial evidence of validity. As described in the results, the correlation between the unsupervised search-based signal with minimised news effects and confirmed COVID-19 cases is maximised when the latter is brought forward by $16.7$ days. This temporal difference between online searches and clinical information could be partly explained by the amount of time between the onset of common symptoms and of more severe ones that warrant a hospital admission~\cite{Zhou2020}, the overall delay of some health systems to respond to the pandemic~\cite{Sohrabi2020}, and national policies that recommended testing only after the persistence of symptoms in milder cases~\cite{hale2020variation}. Interestingly, if no minimisation of news effects is conducted, the temporal distance for a maximal correlation increases by $1.7$ days. The fact that most countries in our analysis exhibited concern about COVID-19 prior to the rapid increase of infections could be a justification for this. By repeating this analysis for mortality figures, we obtain that correlation is maximised by bringing death time series $22.1$ ($17.4-26.9$) days forward. The added amount of days corroborates with findings about the time interval between the commencement of a hospitalisation and a death outcome ($5-11$ days)~\cite{Zhou2020}.

\begin{figure}[t]
    \centering
    \begin{tabular}{cc}
    \multirow{4}{*}[2.5em]{\rotatebox[origin=c]{90}{\sffamily\small Normalised scores}} & \def\stackalignment{l}
    \topinset{\sffamily\textbf{A}}{\includegraphics[height=1in, clip=true, trim=0.3in 0.5in 0 0]{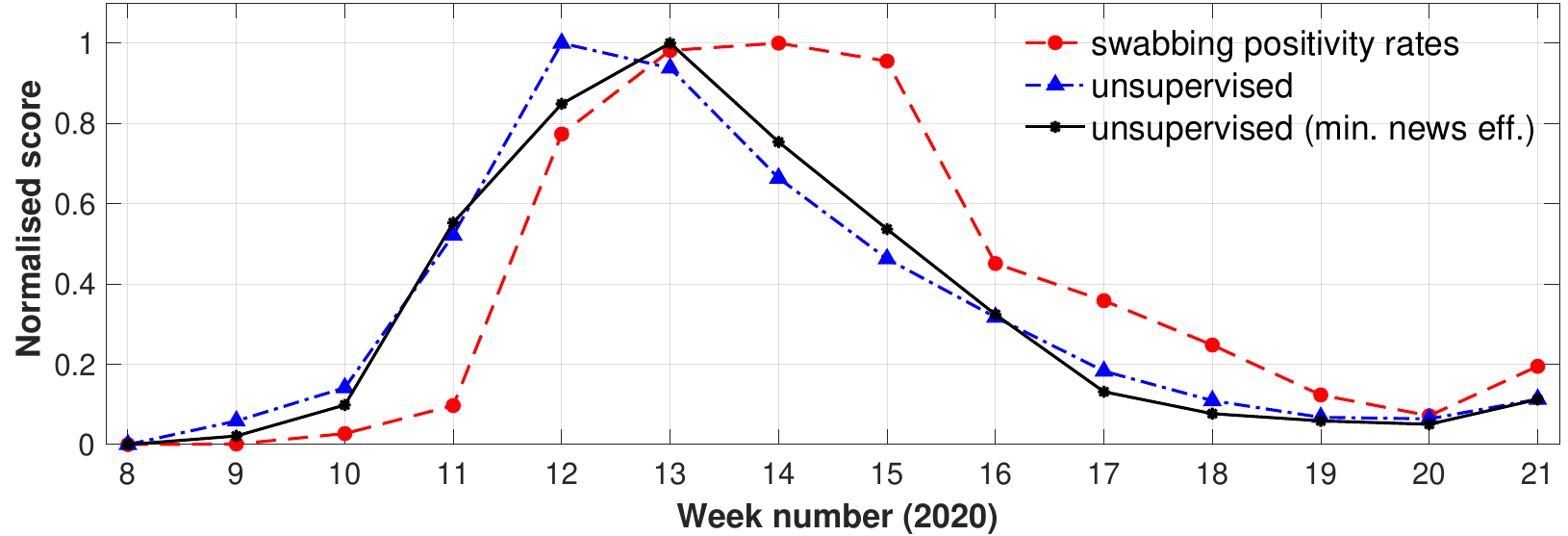}}{0.08in}{0.2in}
    \def\stackalignment{l}
    \topinset{\sffamily\textbf{B}}{\includegraphics[height=1in, clip=true, trim=0.73in 0.5in 0 0]{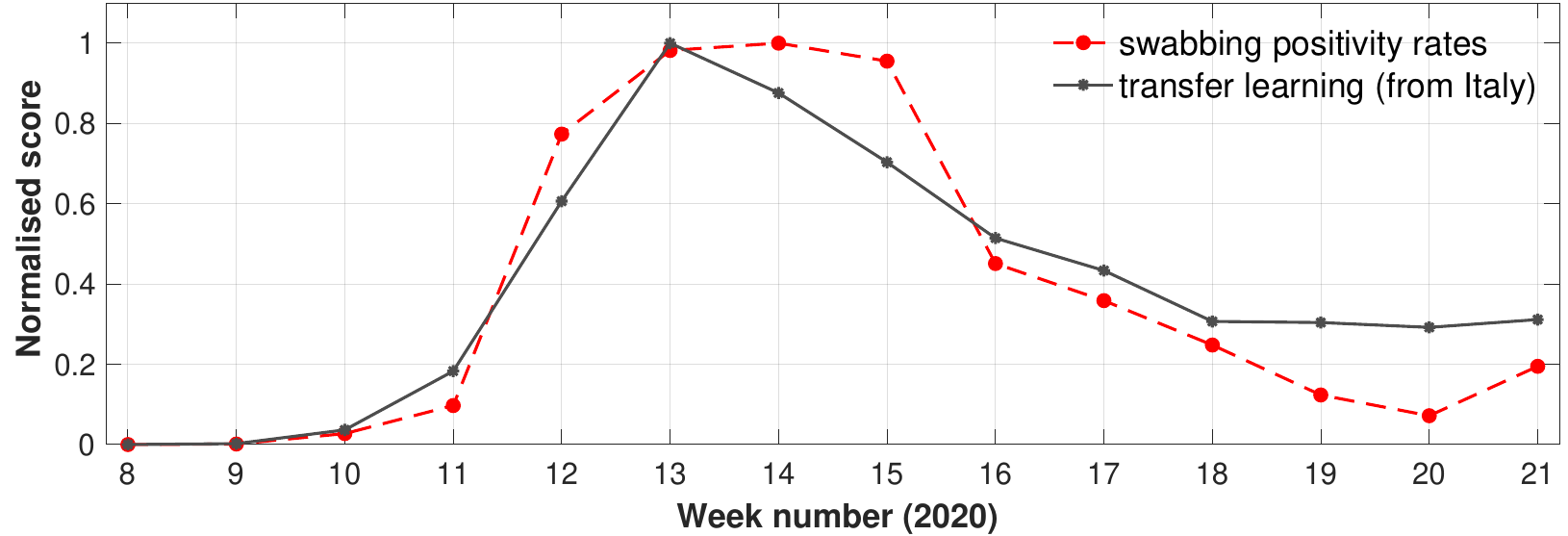}}{0.08in}{0.09in}
    \\
    
    & 
    \def\stackalignment{l}
    \topinset{\sffamily\textbf{C}}{\includegraphics[height=1.0625in, clip=true, trim=0.3in 0.3in 0 0]{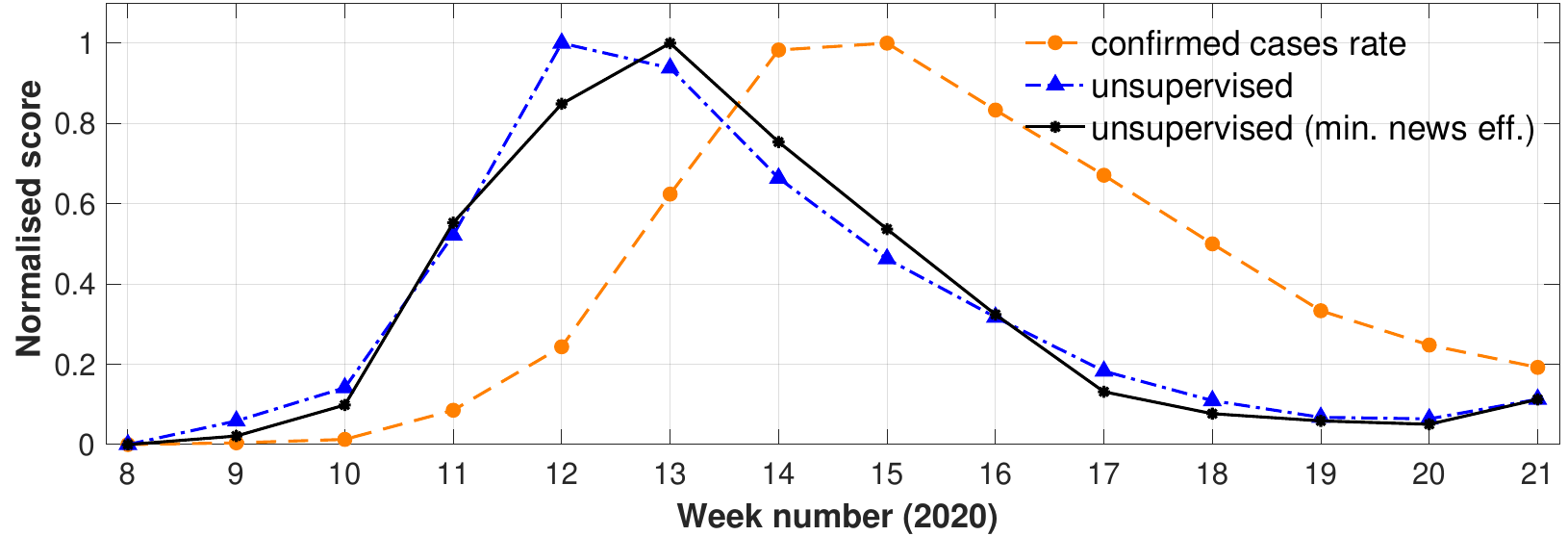}}{0.08in}{0.2in}
    \def\stackalignment{l}
    \topinset{\sffamily\textbf{D}}{\includegraphics[height=1.0625in, clip=true, trim=0.73in 0.3in 0 0]{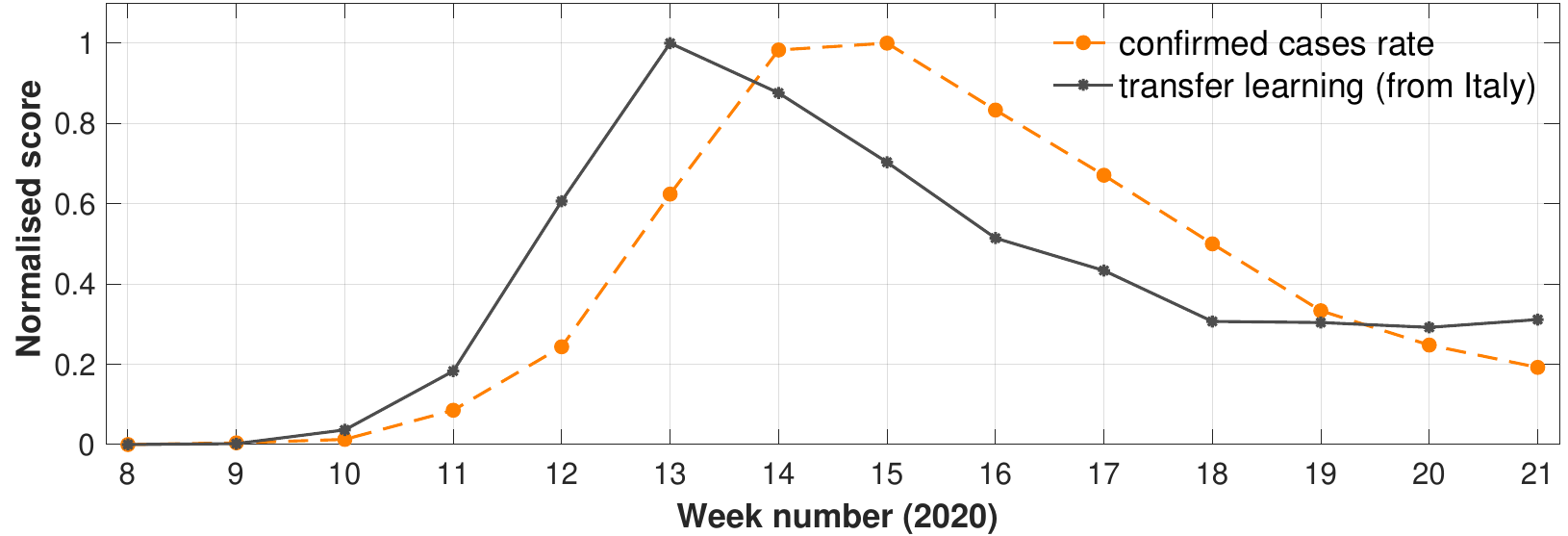}}{0.08in}{0.09in}
    \\
    
    \multicolumn{2}{c}{\sffamily\small\hspace{0.35in} Week number (2020)}
    
    \end{tabular}

    \caption{Comparison of weekly online search-based signals for COVID-19 to different clinical endpoints for England. (\textbf{A})~Estimates from the unsupervised models with or without minimising news media effects are compared to COVID-19 positivity rates obtained through a swabbing test scheme operated by the RCGP. (\textbf{B})~Estimates for COVID-19 obtained via transfer learning (source model of confirmed cases is based on data from Italy) are compared to COVID-19 positivity rates obtained through a swabbing test scheme ran by the RCGP. (\textbf{C})~Estimates from the unsupervised models with or without minimising news media effects are compared to confirmed cases rates as reported by PHE. (\textbf{D})~Estimates for COVID-19 obtained via transfer learning (source model of confirmed cases is based on data from Italy) are compared to confirmed cases rates as reported by PHE.}
    \label{fig:comp_diff_endpoints}
\end{figure}

Drawing our focus to England, we were able to compare outcomes from the unsupervised (with and without minimised effects) and transfer learning approaches to more conventional metrics. The first is a swabbing scheme led by the Royal College of General Practitioners (RCGP), which included many patients who did not have COVID-19-related symptoms and thus allowing us to obtain a more representative metric for community-level spread~\cite{delusignan2020}; results are depicted in Figs.~\ref{fig:comp_diff_endpoints}(A)~and~\ref{fig:comp_diff_endpoints}(B). We observe that the weekly COVID-19 swabbing positivity rates correlate strongly with the unsupervised ($r = .816$, $p < .001$), unsupervised with minimised news media effects ($r = .855$, $p < .001$), and transferred ($r = .954$, $p < .001$) scores. We find that the unsupervised models could have provided an early warning at least a week before the RCGP swabbing scheme. The second metric is a normalised version of the confirmed cases that takes into consideration relative population figures (\href{https://coronavirus.data.gov.uk/}{coronavirus.data.gov.uk}); comparisons are depicted in Figs.~\ref{fig:comp_diff_endpoints}(C)~and~\ref{fig:comp_diff_endpoints}(D). Here, the time discrepancy is more visible, and correlations without any form of shifting are within a similar range as in our previous analysis (see Figs.~\ref{fig:uns_vs_cc}~and~\ref{fig:results_tl}), showing that search-based estimates precede the clinical estimates by 1 and 2 weeks for the transfer learning and unsupervised approaches respectively.

{\sloppy
PHE in the UK has incorporated the unsupervised models described in this paper into their weekly surveillance reports about COVID-19  (\href{https://www.gov.uk/government/publications/national-covid-19-surveillance-reports}{gov.uk/government/publications/national-covid-19-surveillance-reports}).} According to PHE, estimates from these models provided important insights into community transmission at a time where conventional surveillance had limited reach due to the fact that testing was restricted to hospitalised cases. They were analysed alongside other surveillance systems, with the broad expectation that any changes in trends would be first detected through the analysis of web searches and would then be seen at a later stage in more conventional surveillance sources. The decrease in the web search scores was interpreted as an early signal of the impact of physical distancing~\cite{jarvis2020quantifying}. The ability to quantify the lag between search engine trends and incidence trends in confirmed cases and mortality further improves the utility of this approach.

Nonetheless, our study is not without limitations. Primarily, it is difficult to provide a solid evaluation for our findings. This will require a better representation of disease incidence from a clinical or epidemiological perspective, which is an ongoing challenge. In addition, wherever a form of statistical supervision has been introduced (transfer learning, correlation and regression analysis, forecasting), reliable outcomes depend on the existence of a consistent sampling technique of the target information, which in this case is confirmed cases or deaths. We made a concerted effort to reduce the impact of such inconsistencies by training models jointly across multiple countries, focusing on mortality figures rather than the testing-dependent confirmed cases for more demanding tasks, and assessing qualitatively our outcomes across different locations. In addition, more detailed comparisons based on public health data from England reaffirm our broader findings. Similarly to other studies that are based on internet data~\cite{lampos2015gft,yang2015accurate}, our approach is likely to have limited applicability to locations with lower rates of internet access.

\section*{Methods}
We first present the methodological approaches deployed in our analysis, and then describe the data sets used.

\subsection*{Community-level surveillance with an unsupervised symptom-based online search model} 
We generate $k$ symptom-based search query sets using the $k$ identified symptoms from the FF100 survey for COVID-19 ($k = 19$)~\cite{Boddington2020}. We also consider one additional set that includes specific COVID-19 terminology, i.e. the ``COVID-19'' keyword itself among others. Query sets may include different wordings for the same symptom or queries with minor grammatical differences (especially for queries in Greek and French). If a symptom is represented by more than one search query, then we obtain the total frequency (sum) across these queries. Each query set time series is smoothed using a harmonic mean over the past 14 days (see Eq.~\ref{eq:harmonic} for a definition of the harmonic mean), and any trends across the entire period of the analysis are removed using linear detrending. We then apply a min-max normalisation (Eq.~S1) to the frequency time series of each query set to obtain a balanced representation between more and less frequent searches. We divide our data into two periods of interest, the current one (from September 30, 2019 until \analysisDate) and a historical one (from September 30, 2011 to September 29, 2019). The corresponding data sets are denoted by $\mathbf{X} \in \mathbb{R}_{\ge 0}^{N_1 \times k}$ and $\mathbf{H} \in \mathbb{R}_{\ge 0}^{N_2 \times k}$, where $N_1$, $N_2$ represent the different numbers of days in the current and historical data, respectively. We use the symptom conditional probability distribution from the FF100 to assign weights ($\mathbf{w} \in \mathbb{R}_{\ge 0}^{k}$) to each query category, and compute weighted time series ($\mathbf{x} = \mathbf{Xw}$, $\mathbf{h} = \mathbf{Hw}$), which are subsequently divided by the sum of $\mathbf{w}$ (weighted average). The additional set that includes specific COVID-19 terms is assigned a weight of $1$. Our rationale is that people who experience COVID-19-related symptoms in addition to searching about them, will also issue queries about the disease as it is a broadly discussed topic. For the historical data, we divide their time span into yearly periods, and compute an average time series trend, $\mathbf{h}_{\mu}$, using two standard deviations as upper and lower confidence intervals.

\subsection*{Minimising the effect of news media using autoregression}
On any given day the proportion of news articles about the COVID-19 pandemic is $m \in [0,1]$, and the weighted score of symptom-related online searches (see previous paragraph) is equal to $g$; we can apply a min-max normalisation so that $g \in [0,1]$ as well. We hypothesise that $g$ incorporates two signals based on infected ($g_p$) and concerned ($g_c$) users, respectively, i.e. 
\begin{equation}
g = g_p + g_c \, .
\end{equation}
Then, there exists a constant $\gamma \in [0,1]$ such that \begin{equation}
g_p = \gamma g \, \, \text{ and } \, \, g_c = (1-\gamma) g \, .
\end{equation}
Our goal is to approximate $\gamma$ using the observed variables $m$ and $g$. The rationale of the deployed approach is similar to the logic behind a Granger causality test~\cite{granger1969}. First, we train a linear AR model for forecasting the online search score at a time point (day) $t$, $g_t$, using its previous values; this is denoted by AR$\left(g\right)$. We also train a linear AR model with the same forecasting target, but an expanded space of observations that includes current ($m_t$) and previous (e.g. $m_{t-1}$) values of the news articles ratio; this is denoted by AR$\left(g,m\right)$. We then use the error ratio between the two models in forecasting $g_t$ as an estimate of $\gamma$ for time point $t$. In particular, we first solve 
\begin{equation}
\label{eq:news_db_ar}
\argmin_{\mathbf{w},b_1} \frac{1}{N}\sum_{t=1}^{N} (g_t - w_1 g_{t-1} - w_2 g_{t-2} - b_1)^2 \, ,
\end{equation}
to learn a pair of weights ($\mathbf{w}$) and an intercept term ($b_1$) for AR$\left(g\right)$. We use 2 lags (past values) to keep the complexity of the task tractable given the small amount of samples at our disposal, $N$. In our experiments we use the previous $N = 56$ days, an intermediate value to account for the trade-off between the amount of samples and maintaining recency. We then solve
\begin{equation}
\label{eq:news_db_ar2}
\argmin_{\mathbf{w},\mathbf{v},b_2} \frac{1}{N}\sum_{t=1}^{N} (g_t - w_1 g_{t-1} - w_2 g_{t-2} - v_1 m_t - v_2 m_{t-1} - v_3 m_{t-2} - b_2)^2 \, ,
\end{equation}
to learn the weights ($[\mathbf{w}; \mathbf{v}]$) and an intercept term ($b_2$) for AR$\left(g,m\right)$. Using both models, we forecast the next (unseen) value of $g$, which is denoted by $\hat{g}_{t+1}$, and compute the absolute error from its known true value, $g_{t+1}$. This yields errors, $\epsilon_1$ and $\epsilon_2$ for AR$\left(g\right)$ and AR$\left(g,m\right)$, respectively. If $\epsilon_1 < \epsilon_2$, then the news media signal does not help to improve the accuracy of AR$\left(g\right)$, and hence we assume that it does not affect the online searches. Otherwise, we estimate its effect to be represented by 
\begin{equation}
\label{eq:news_db_ar_final}
\gamma = \frac{\epsilon_2}{\epsilon_1} \, .
\end{equation}
After obtaining a time series of $\gamma$'s for the all days in our analysis, we smooth each one of them using a harmonic mean (see Eq.~\ref{eq:harmonic}) over the values of the previous 6 days (or 7 days including the day of focus).

\subsection*{Transferring supervised models of confirmed COVID-19 cases to different countries}
Previous work has shown that it is possible to transfer a model for seasonal flu, based on online search query frequency time series, from one country that has access to historical syndromic surveillance data to another that has not~\cite{zou2019www}. Here, we adapt this method to transfer a model for COVID-19 incidence from a source country where the disease spread has progressed significantly to a target country that is still in earlier stages of the epidemic. The main motivation for this is our assumption that a supervised model based on data from the source country might be able to capture disease dynamics sooner and therefore better than unsupervised or supervised models based solely on data from the target countries. The steps and data transformations that are required to apply this technique are detailed below.

Search query frequency time series are denoted by $\mathbf{S} \in \mathbb{R}^{M \times n_\text{S}}_{\ge 0}$ and $\mathbf{T} \in \mathbb{R}^{M \times n_\text{T}}_{\ge 0}$, for the source and target countries respectively; $M$ denotes the number of days considered, and $n_\text{S}$, $n_\text{T}$ the number of queries for the two locations. As these time series are quite volatile for some locations in our study, something that does not help in cross-location mapping of the data, we have smoothed them using a harmonic query frequency mean based on a window of $D = 14$ past days (see Eq.~\ref{eq:harmonic}). We subsequently train an elastic net model on data from the source location~\cite{zou2005elastic}, similarly to previous work on ILI~\cite{lampos2015gft,lampos2015assessing,lampos2017www} or other text regression tasks~\cite{lampos2013user,lampos2014predicting}. In particular, we solve the following optimisation task
\begin{equation}
    \label{eq:elastic_net}
    \argmin_{\mathbf{w},b} \left (\left \lVert \mathbf{y} - \mathbf{S} \mathbf{w} - b \right \rVert_{2}^{2} + \lambda_1 \left \lVert \mathbf{w} \right \rVert_{1} + \lambda_2 \left \lVert \mathbf{w} \right \rVert_{2}^{2}        \right ) \, ,
\end{equation}
where $\mathbf{y} \in \mathbb{R}^M$ denotes the daily number of confirmed COVID-19 cases in the source location, $\lambda_1$, $\lambda_2 \in \mathbb{R}_{>0}$ are the $\ell_1$- and $\ell_2$-norm regularisation parameters, and $\mathbf{w} \in \mathbb{R}^{n_\text{S}}$, $b \in \mathbb{R}$ denote the query weights and regression intercept, respectively. Prior to deploying elastic net, we apply a min-max normalisation on both $\mathbf{S}$ and $\mathbf{y}$. We fix the ratio of $\lambda$'s, and then train $q$ models for different values of $\lambda_1$ starting from the largest value that does not yield a null model and exploring the entire regularisation path. From all the different regression models represented by the columns of $\mathbf{W} \in \mathbb{R}^{n_\text{S} \times q}$, and the elements of $\mathbf{b} \in \mathbb{R}^q$, we use the ones that satisfy a sparsity level from $5.56\%$ to $90.74\%$ ($3$ to $49$ search queries out of a total of $54$) as an ensemble to accomplish a more inclusive transfer.

To generate an equivalent feature space for the target location (same dimensionality, similar feature attributes), we first establish query set pairs between the source and the target location using the symptom categories in the FF100 survey. We map a source query to the target query from the same symptom category that maximises their linear correlation based on their frequency time series. As the time series of source and target queries may not be temporally aligned, prior to computing correlations, we look at a window of $z = 45$ days (backwards and forwards) and identify at which temporal shift the average correlation between search query frequencies in $\mathbf{S}'$ and $\mathbf{T}$ is maximised; $\mathbf{S}'$ here denotes a subset of $\mathbf{S}$ that includes only the search queries that have been assigned a non zero weight by the elastic net (Eq.~\ref{eq:elastic_net}). In the rare event that no active target search query exists for a certain symptom category, to maintain model consistency we use the best correlated one from all target queries available (irrespectively of the symptom category) as its mapping. After this process, we end up with a subset $\mathbf{Z} \in \mathbf{R}^{M \times n_\text{S}}$ of the target feature space $\mathbf{T}$. Notably, $\mathbf{Z}$ does not necessarily hold data for $n_S$ distinct queries as different source queries may have been mapped to the same target query. $\mathbf{Z}$ is subsequently normalised using min-max. To reduce the distance in the query frequency range between the two feature spaces ($\mathbf{S}$, $\mathbf{Z}$), we scale the latter based on their mean, column-wise (per search query) ratio $\mathbf{r} \in \mathbb{R}^{n_\text{S}}_{\ge 0}$, i.e. $\mathbf{Z}_S = \mathbf{Z} \odot \mathbf{r}$. We then deploy the ensemble source models to the target space, making multiple inferences (for different $\lambda_1$ values) held in $\mathbf{Y} \in \mathbb{R}^{n_\text{S} \times q}$:
\begin{equation}
    \mathbf{Y} = \mathbf{Z}_S \mathbf{W} + \mathbf{b} \, .
\end{equation}
We reverse the min-max normalisation for each one of the inferred time series (columns of $\mathbf{Y}$) using values from the source model's ground truth $\mathbf{y}$ (prior to its normalisation). Finally, we compute the mean of the ensemble (across the rows of $\mathbf{Y}$) as our target estimate, and also use the $.025$ and $.975$ quantiles to form $95$\% confidence intervals.

\subsection*{Correlation and regression analysis}
The relationship of search frequency time series and confirmed cases or deaths can uncover symptoms or behaviours related to COVID-19. However, as it is hard to find resources that are representative of community level disease rates, looking at this relationship separately for each country might produce misleading outcomes. To mitigate this to the extent possible, we combine data from $C$ countries and produce an aggregate set of query frequencies, $\mathbf{Z}_{\aggsym} \in \mathbb{R}^{CM \times n}$, where $M$, $n$ denote the considered days and search queries, respectively. We denote the aggregated daily confirmed COVID-19 cases or deaths for these countries with $\mathbf{y}_{\aggsym} \in \mathbb{R}^{CM}$. Prior to the aggregation, we apply min-max normalisation on the query frequency, confirmed cases, and deaths time series separately for each country to balance out local properties. Initially, we compute the linear correlation between the columns of $\mathbf{Z}_{\aggsym}$ and $\mathbf{y}_{\aggsym}$. Correlation is an informative metric, but considers each search query in isolation. Therefore, we also perform a multivariate regression analysis to more rigorously estimate the impact of each search query in estimating $\mathbf{y}_{\aggsym}$. To do this, we apply elastic net regularised regression (see Eq.~\ref{eq:elastic_net}), training and testing $K$ models, one for each day of an identified test period of interest. During each of the $K$ training phases, assuming it contains $\eta$ days, we use data up to and including the past $\eta-1$ days to train, and test only on the last day ($\eta_{\text{th}}$) which is unseen; this results in daily test sets of size $C$ (one value for each country). We explore elastic net's regularisation path to consider $L$ models that maintain (by assigning a nonzero weight) up to a reasonable percentage of the features (e.g. $50\%$), so that a solution is not overfitting. We do this gradually, selecting first $1\%$ of the features and moving towards the maximum considered percentage. In this experiment, we use the test set to identify the most accurate (in terms of mean squared error) model at each density level. For this model, we determine the impact of each one of the features (search queries) by considering both its frequency and allocated weight. The impact $\Theta(\cdot)$ of a query $q$ is equal to
\begin{equation}
    \Theta(q) = \left(\sum_{\ell=1}^{L} \sum_{t=1}^{K} \sum_{j=1}^{C} f_{t,j} \, \, w_{\ell,t}\right) \bigg/ \left(\sum_{\ell=1}^{L} \sum_{t=1}^{K}\sum_{j=1}^{C} \hat{y}_{\ell,t,j} \right) \, , 
\end{equation}
where $f_{t,j}$ denotes the query frequency at time point (day) $t$ and for country $j$, $w_{\ell,t}$ the corresponding weight at sparsity level $\ell$, and $\hat{y}_{\ell,t,j}$ the respective estimated confirmed cases or deaths. Impacts are summed across all the considered days, and model densities, and normalised at the end by the sum of all the corresponding estimates.

\subsection*{Short-term forecasting of COVID-19 deaths}
Moving further into supervised models, we develop forecasting solutions for COVID-19 to assess the potential value of incorporating web search data in AR models. In this case, we focus on modelling related deaths as opposed to confirmed cases because the reporting of the latter has been dependent on testing approaches, and thus has a less reliable time series structure from a statistical perspective. Short-term forecasts of COVID-19 deaths can be conducted using the time series of past records. Augmenting this AR signal with online user trails could help to improve accuracy, if we draw a parallel with ILI rate modelling~\cite{paul2014twitter}. Let $\mathbf{z} \in \mathbb{R}_{\ge 0}^{M}$ denote the average frequency of $N$ min-max normalised search queries for $M$ days, and let $\mathbf{y} \in \mathbb{N}^{M}$ be the corresponding COVID-19 deaths. We first solve a strictly AR task using $L$ past values, meaning that at a time point $t$ we use $\mathbf{y}_{\text{AR}}(t,L) = [y_t, y_{t-1}, \dots, y_{t-L}] \in \mathbf{y}$ to forecast $y_{t+D}$, performing $D$ days ahead forecasting. We denote this forecasting model as AR-F. We also augment our observations by incorporating search query frequency data for the same time window resulting to an input $\mathbf{x}_{\text{AR}}(t,L) = [\mathbf{z}_{\text{AR}}(t,L); \mathbf{y}_{\text{AR}}(t,L)]$ that is held in the concatenated matrix $\mathbf{X} = [\mathbf{Z}_{\text{AR}}(L); \mathbf{Y}_{\text{AR}}(L)] \in \mathbb{R}_{\ge 0}^{M \times 2(L+1)}$. We denote this search AR forecasting model as SAR-F. For simplicity, we drop the subscript notation of the input variables ($\mathbf{x}$, $\mathbf{z}$, $\mathbf{y}$) in subsequent references to them.

We develop forecasting models using Gaussian Processes (GP), training a different model for each $D$ days ahead forecasting task. The choice of GPs is justified by previous work on modelling infectious diseases~\cite{lampos2015gft,lampos2017www,zou2018multi}. GPs are defined as random variables any finite number of which have a multivariate Gaussian distribution \cite{rasmussen2006}. GP methods aim to learn a function $f : \mathbb{R}^m \rightarrow \mathbb{R}$ drawn from a GP prior. They are specified through a mean and a covariance (or kernel) function, i.e. $f(\mathbf{x}) \sim \text{GP}(\mu(\mathbf{x}),k(\mathbf{x},\mathbf{x}'))$, where $\mathbf{x}$ and $\mathbf{x}'$ (both $\in \mathbb{R}_{\ge 0}^{2(L+1)}$) denote rows of the input matrix $\mathbf{X}$ for the SAR-F model. By setting $\mu(\mathbf{x}) = 0$, a common practice in GP modelling, we focus only on the kernel function. The specific kernel function used in SAR-F is given by
\begin{equation}
\label{eq:gp_kernel}
    k(\mathbf{x},\mathbf{x}') = k_\text{SE}(\mathbf{z},\mathbf{z}';\sigma_1,\ell_1) + k_\text{SE}(\mathbf{y},\mathbf{y}';\sigma_2,\ell_2) +
    k_\text{SE}(\mathbf{x},\mathbf{x}';\sigma_3,\ell_3) +
    \sigma^2_4\delta(\mathbf{x},\mathbf{x}') \, ,
\end{equation}
where $k_\text{SE}(\cdot,\cdot)$ denotes a squared exponential (SE) covariance function, $\delta(\cdot,\cdot)$ denotes a Kronecker delta function used for an independent noise component, $\ell$'s and $\sigma$'s are lengthscale and scaling (variance) parameters, respectively. This composite covariance function uses separate kernels for online search and deaths data, as well as a kernel where they are modelled jointly. This provides the GP with more flexibility in adapting its decisions to one class of observations over the other. For completeness, the SE kernel is defined by
\begin{equation}
k_\text{SE}(\mathbf{x},\mathbf{x}') = \sigma^2 \exp{\left(- \frac{(\mathbf{x}-\mathbf{x}')^{2}}{2\ell^{2}}\right)} \, .
\end{equation}
The covariance function of the AR-F model is a simplified version of Eq.~\ref{eq:gp_kernel}, where search data is not used, i.e.
\begin{equation}
    k(\mathbf{y},\mathbf{y}') =  k_\text{SE}(\mathbf{y},\mathbf{y}';\sigma_1,\ell_1) +
    k_\text{SE}(\mathbf{y},\mathbf{y}';\sigma_2,\ell_2) +
    \sigma^2_3\delta(\mathbf{y},\mathbf{y}') \, .
\end{equation}
The hyperparameters of both covariance functions ($\sigma$'s, $\ell$'s) are optimised using Gaussian likelihood and variational inference~\cite{rasmussen2006}.

In our analysis, we compare the aforementioned forecasting models with a basic persistence model (PER-F). For a $D$ days ahead forecasting task, PER-F uses the most recently observed value of the ground truth, $y_t$, as the forecasting estimate for the time instance $t+D$ ($\hat{y}_{t+D} = y_t$). Given the limited time span and the irregularities in reporting, this is a competitive baseline to improve upon.

\subsection*{Data sets}
Online search data is obtained from Google Health Trends, a non public application interface offered by Google for research on health-related topics. Data represent daily online search query frequencies for specific areas of interest. Query frequencies are defined as the sum of search sessions that include a target search term divided by the total number of search sessions (for a day and area of interest). Hence, in the paper when we refer to a certain search query, we also refer to all other search queries that include all of its words. Google defines a search session as a grouping of consecutive searches by the same user within a short time interval. We have obtained data from September 30, 2011 to \analysisDate~for the US, UK, England, Australia, Canada, France, Italy, Greece, and South Africa. The list of search terms is determined by COVID-related symptoms and keywords. For each country, we mainly used queries in its native language(s). Each daily search query frequency is smoothed using a harmonic mean over the past $D$ days (including itself). More specifically, a smoothed search query frequency $s_{i}$ for a day $i$ is equal to
\begin{equation}
\label{eq:harmonic}
    s_i = \frac{1}{\sum_{p=1}^{D}\frac{1}{p}}\sum_{p=1}^{D} \frac{x_{i-p+1}}{p} \, ,
\end{equation}
where $x_{(\cdot)}$ denotes the raw (non smoothed) frequency. In our experiments we set $D = 14$ days.

We are also using a global news corpus to extract media coverage trends for COVID-19 in all the countries in our study. This is estimated by counting the proportion of articles mentioning a COVID-19-related term (see SI for the list of terms). In particular, daily counts of the total of news media articles published, and the subset that included at least one relevant keyword anywhere in the body of the text were collected from the Media Cloud database (\href{https://mediacloud.org/}{mediacloud.org}) for the UK ($93$), US ($225$), Australia ($61$), Canada ($79$), France ($360$), Italy ($178$), Greece ($75$), and South Africa ($135$), where in the parentheses we state the number of media sources considered per country. These counts were collected from September 30, 2019 through \analysisDate. Within this time-span we identified $2{,}535{,}735$ articles as COVID-19-related from a total of $10{,}093{,}349$. Fig.~S9 depicts the average daily ratio across all countries, as soon as it started being above zero (Jan. 2020), with two standard deviations as confidence intervals. Initially, there exists a distinctive pattern of (exponential) increase, but from April onwards the average media coverage has a moderate decreasing trend. In addition, we observe a certain variance across locations or time periods, that adds to the potential value of this signal.

For all countries in our analysis, we obtained daily confirmed COVID-19 cases and deaths time series from the European Centre for Disease Prevention and Control (ECDC). Links are provided in the SI.

Finally, we used data from the NHS first few hundred (FF100) survey based on people who have contracted SARS-CoV-2~\cite{Boddington2020}. The FF100 survey has identified $19$ symptoms that could be caused by COVID-19 with the following probabilities in decreasing order: cough ($.777$), fatigue ($.709$), fever ($.601$), headache ($.567$), muscle ache ($.509$), appetite loss ($.441$), shortness of breath ($.404$), sore throat ($.386$), joint ache ($.339$), runny nose ($.325$), loss of the sense of smell ($.291$), diarrhoea ($.276$), sneezing ($.239$), nausea ($.236$), vomiting ($.087$), altered consciousness ($.068$), nose bleed ($.060$), rash ($.052$), and seizure ($.008$).

\subsection*{Ethics}
Ethical approval was not required for the analysis presented in this paper. Data was obtained in an anonymised and aggregated (national level) format.

\section*{Data availability}
The online search and RCGP swabbing scheme data sets that support the findings of this study are available from Google and RCGP/PHE, respectively. Restrictions apply to the availability of these data sets, which were used under license for the current study, and so are not publicly available. These data sets are however available from the authors upon reasonable request and with the respective permission of Google and RCGP/PHE. The rest of the data sets that this study is based on are available at \href{https://figshare.com/projects/Tracking\_COVID-19\_using\_online\_search/81548}{figshare.com/projects/Tracking\_COVID-19\_using\_online\_search/81548}.

\bibliography{refs}

\section*{Acknowledgements}
V.L, S.M., R.A.M., and I.J.C would like to acknowledge all levels of support from the EPSRC projects ``EPSRC IRC in Early-Warning Sensing Systems for Infectious Diseases'' (EP/K031953/1) and ``i-sense: EPSRC IRC in Agile Early Warning Sensing Systems for Infectious Diseases and Antimicrobial Resistance'' (EP/R00529X/1). V.L. and I.J.C. would also like to acknowledge the support from the MRC/NIHR project ``COVID-19 Virus Watch: Understanding community incidence, symptom profiles, and transmission of COVID-19 in relation to population movement and behaviour'' (MC\_PC\_19070). The authors would like to thank the NHS/PHE FF100 team for sharing early results, and the RCGP for sharing swabbing data for COVID-19. We also appreciate the contribution of Ettore Severi, Anna Odone, and Daniela Paolotti in the translation of search queries from English to Italian. Finally, V.L. would like to thank Sam J. Gilbert for interesting discussions and pointers during the development of this work.

\section*{Author contribution statement}
V.L. conceived this research project, formed the majority of the data sets, conceived and developed the methods, ran the experiments, wrote the first drafts of this manuscript, and led the write-up thereafter. M.S.M. provided news coverage data for all countries in our analysis. M.E. provided a translation of search query groups from English to Italian, S.M. from English to French, and M.X.R. and Y.H. from English to various languages spoken in South Africa. M.S.M., E.Y.-T., M.E., S.M., Y.H., and I.J.C. contributed in the write-up of the manuscript. All authors provided feedback in various levels of this work.

\newpage

\newcommand{\beginsupplement}{%
        \setcounter{table}{0}
        \renewcommand{\thetable}{S\arabic{table}}%
        \setcounter{figure}{0}
        \renewcommand{\thefigure}{S\arabic{figure}}%
        \setcounter{equation}{0}
        \renewcommand{\theequation}{S\arabic{equation}}%
}

\thispagestyle{empty}
\beginsupplement

\begin{center}
{
\sffamily\LARGE\bfseries Supplementary Information
}
\end{center}

\vspace{0.4in}

\section*{Additional information about the data sets}
\sloppy
Confirmed COVID-19 cases and deaths were obtained from the European Centre for Disease Prevention and Control (ECDC) at \href{https://www.ecdc.europa.eu/en/publications-data/download-todays-data-geographic-distribution-covid-19-cases-worldwide}{ecdc.europa.eu/en/publications-data/download-todays-data-geographic-distribution-covid-19-cases-worldwide}. All search queries that were considered in our experiments are available online at \href{https://figshare.com/projects/Tracking\_COVID-19\_using\_online\_search/81548}{figshare.com/projects/Tracking\_COVID-19\_using\_online\_search/81548}. News articles were considered as COVID-19-related if their title or main text included one of the terms
\begin{itemize}
    \item 
    \textgreek{`κορονοϊός', `κορονοϊού', `κορωνοϊός', `κορωνοϊού', `κορωνοϊοί', `κορονοϊοί'}, `covid', `covid-19', `covid 19', `covid19', `coronavirus', and `ncov' for Greece, or
    \item `covid', `covid-19', `covid 19', `covid19', `coronavirus', `ncov' for the rest of the countries.
\end{itemize}

\section*{Supplementary equations}
The min-max normalisation of a vector $\mathbf{x} \in \mathbb{R}^n$ will result to a vector $\hat{\mathbf{x}} \in [0,1]^n$ by performing the following operation:
\begin{equation}
    \hat{\mathbf{x}} = \frac{\mathbf{x} - \min(\mathbf{x})}{\max(\mathbf{x}) - \min(\mathbf{x})} \, .
\end{equation}
%
The z-score normalisation (or standardisation) of a vector $\mathbf{x} \in \mathbb{R}^n$ will result to a vector $\hat{\mathbf{x}} \in \mathbb{R}^n$ by performing the following operation:
\begin{equation}
    \hat{\mathbf{x}} = \frac{\mathbf{x} - \mu(\mathbf{x})}{\sigma(\mathbf{x})} \, ,
\end{equation}
where $\mu(\cdot)$ and $\sigma(\cdot)$ denote the mean and standard deviation functions respectively.

\noindent The mean absolute error (MAE) between an estimated time series $\hat{\mathbf{y}} \in \mathbb{R}^n$ and the corresponding ground truth $\mathbf{y} \in \mathbb{R}^n$ is equal to
\begin{equation}
    \text{MAE}(\hat{\mathbf{y}},\mathbf{y}) = \frac{1}{n} \sum_{i=1}^n\left|\hat{y_i} - y_i\right| \, .
\end{equation}

\section*{Supplementary figures and tables}
\begin{figure}[ht]
    \centering
    \begin{tabular}{cc}
    \multirow{1}{*}[7.25em]{\rotatebox[origin=c]{90}{\sffamily\small Frequency}} & \def\stackalignment{l}
    \topinset{}{\includegraphics[height=1.54in, clip=true, trim=0.23in 0.2in 0 0]{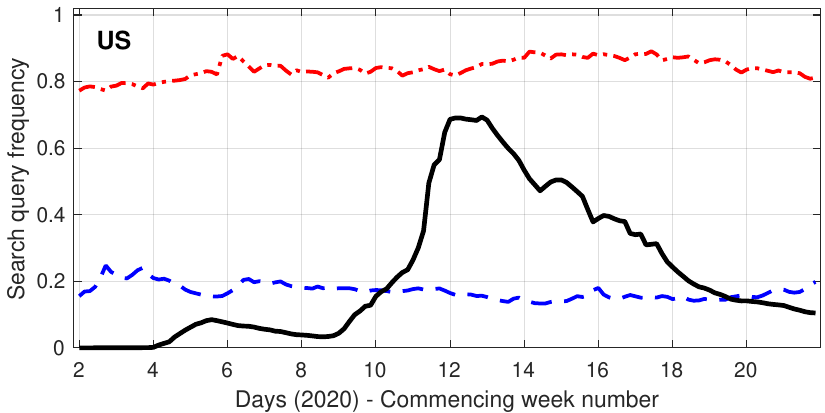}}{0in}{0in}
    \def\stackalignment{l}
    \topinset{}{\hspace{0.04in}\includegraphics[height=1.54in, clip=true, trim=0.45in 0.2in 0 0]{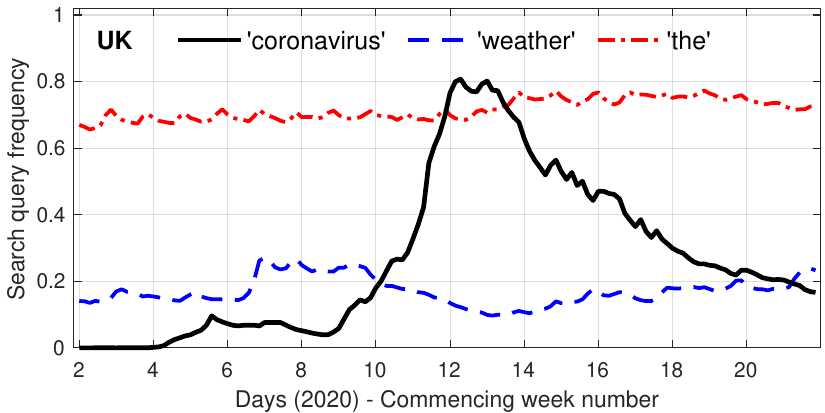}}{0in}{0in} \\
    \multicolumn{2}{c}{\sffamily\small Days (2020) -- Commencing week number}
    \end{tabular}
    \caption{Normalised daily frequency time series of all Google search queries that include the keywords ``coronavirus'', ``weather'', or ``the'' in the US and the UK. Time series are smoothed using a 14-day harmonic mean.}
    \label{fig:searchVol}
\end{figure}

\begin{figure}[ht]
    \centering
    \begin{tabular}{cc}
    \multirow{4}{*}[-1.1em]{\rotatebox[origin=c]{90}{\sffamily\small Normalised online search score for COVID-19}} & \def\stackalignment{l}
    \topinset{}{\includegraphics[height=1.2in, clip=true, trim=0.23in 0.35in 0 0]{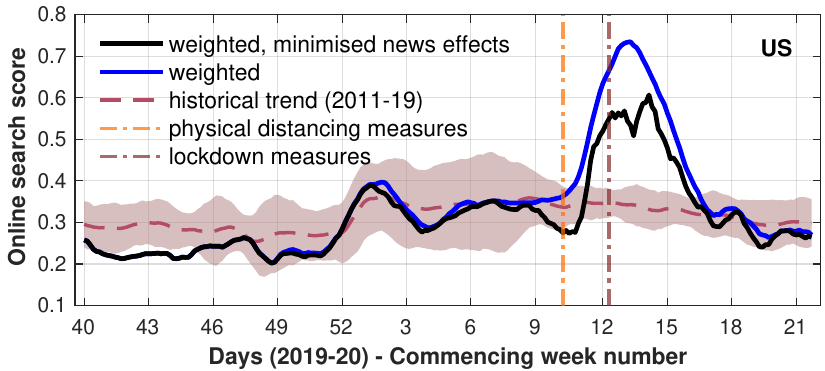}}{0in}{0in}
    \def\stackalignment{l}
    \topinset{}{\hspace{0.04in}\includegraphics[height=1.2in, clip=true, trim=0.45in 0.35in 0 0]{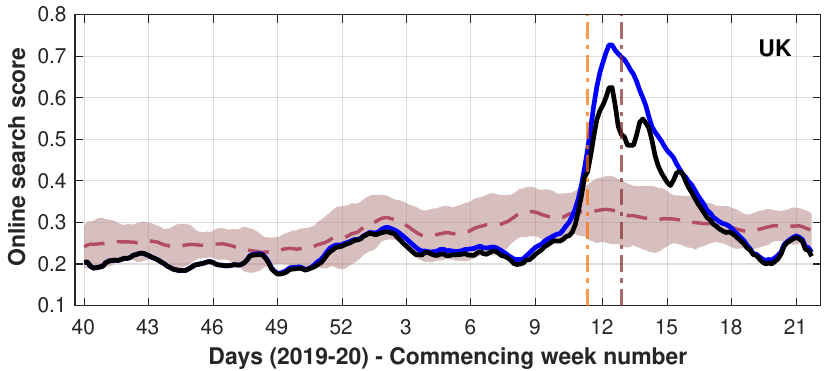}}{0in}{0in} \\
    
    & \def\stackalignment{l}
    \topinset{}{\includegraphics[height=1.2in, clip=true, trim=0.23in 0.35in 0 0]{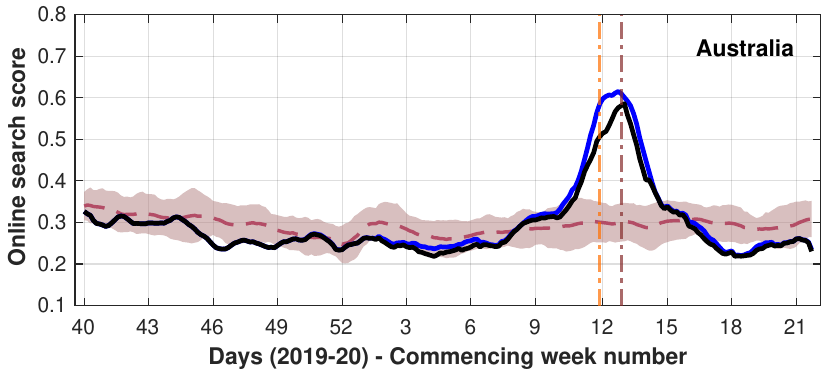}}{0in}{0in}
    \def\stackalignment{l}
    \topinset{}{\hspace{0.04in}\includegraphics[height=1.2in, clip=true, trim=0.45in 0.35in 0 0]{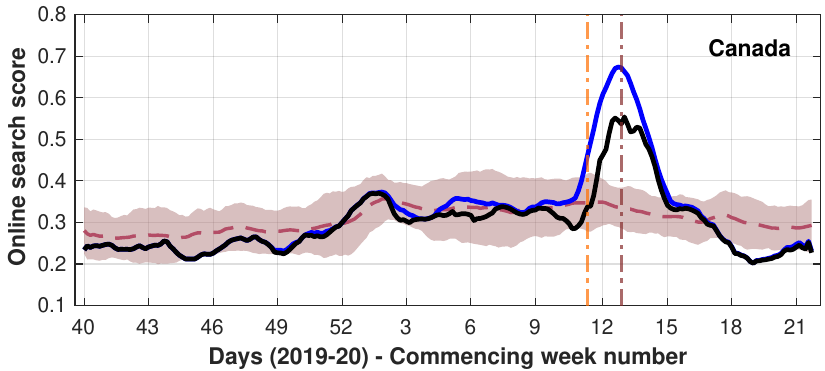}}{0in}{0in} \\
    
    & \def\stackalignment{l}
    \topinset{}{\includegraphics[height=1.2in, clip=true, trim=0.23in 0.35in 0 0]{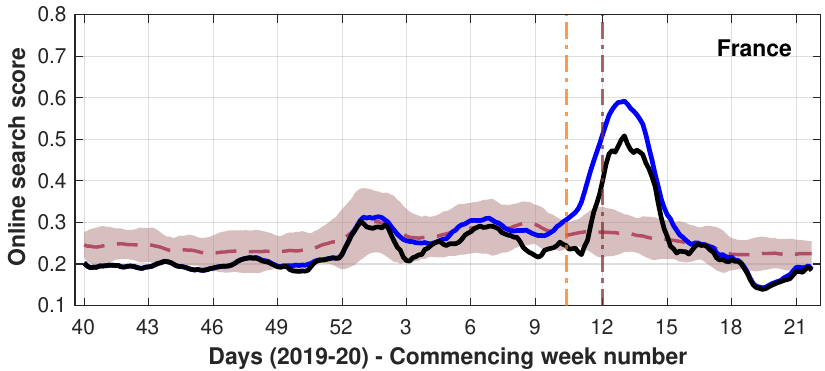}}{0in}{0in}
    \def\stackalignment{l}
    \topinset{}{\hspace{0.04in}\includegraphics[height=1.2in, clip=true, trim=0.45in 0.35in 0 0]{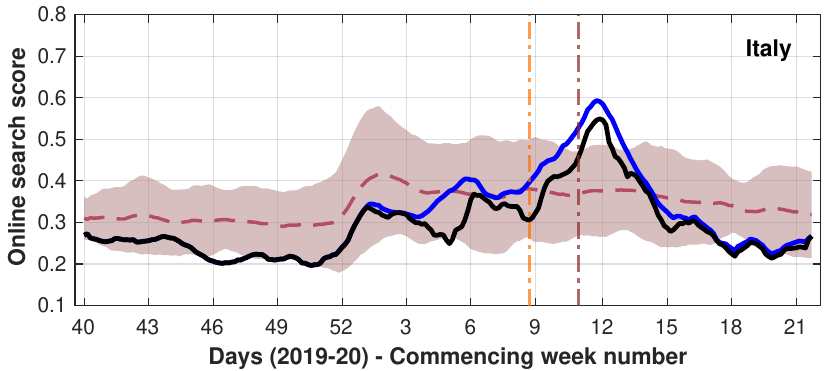}}{0in}{0in} \\
    
    & \def\stackalignment{l}
    \topinset{}{\includegraphics[height=1.285in, clip=true, trim=0.23in 0.20in 0 0]{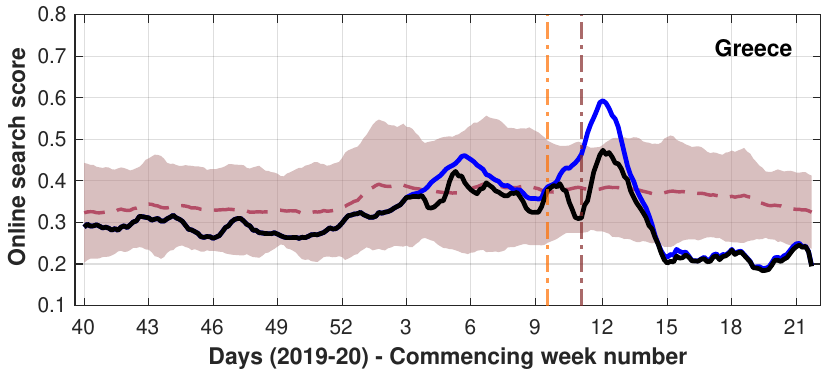}}{0in}{0in}
    \def\stackalignment{l}
    \topinset{}{\hspace{0.04in}\includegraphics[height=1.285in, clip=true, trim=0.45in 0.20in 0 0]{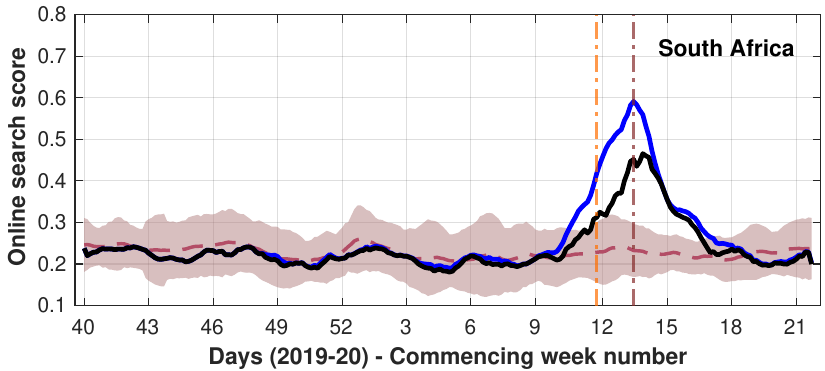}}{0in}{0in}\\
    
    \multicolumn{2}{c}{\sffamily\small\hspace{0.3in} Days (2019-20) -- Commencing week number}
    
    \end{tabular}
    
    \caption{Online search based scores for COVID-19-related symptoms as identified by the FF100 survey for $8$ countries from September 30, 2019 to \analysisDate~(all inclusive). Query frequencies are weighted by symptom occurrence probability (blue line) and have news media effects minimised (black line). These scores are compared to an average $8$-year trend of the weighted model (dashed line) and its corresponding $95\%$ confidence intervals (shaded area). Application dates for physical distancing or lockdown measures are indicated with dash-dotted vertical lines; for countries that deployed different regional approaches, the first application of such measures is depicted. All time series are smoothed using a $7$-point moving average, centred around each day.}
    \label{fig:results}
\end{figure}

\begin{figure}[ht]
    \centering
    \begin{tabular}{cc}
    \multirow{4}{*}[-1.1em]{\rotatebox[origin=c]{90}{\sffamily\small Normalised online search score for COVID-19}} & \def\stackalignment{l}
    \topinset{}{\includegraphics[height=1.2in, clip=true, trim=0.23in 0.35in 0 0]{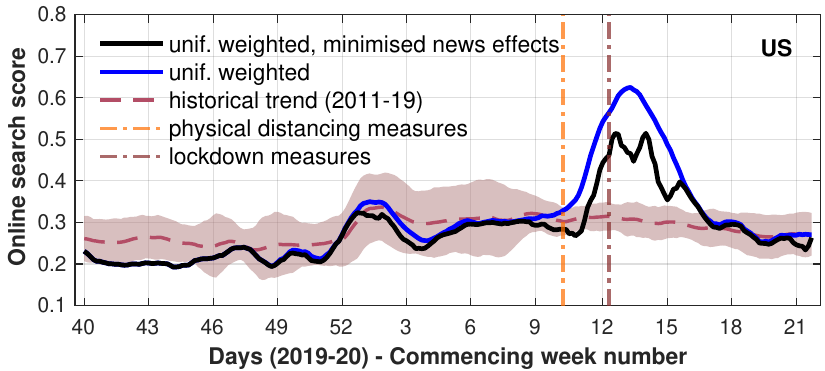}}{0in}{0in}
    \def\stackalignment{l}
    \topinset{}{\hspace{0.04in}\includegraphics[height=1.2in, clip=true, trim=0.45in 0.35in 0 0]{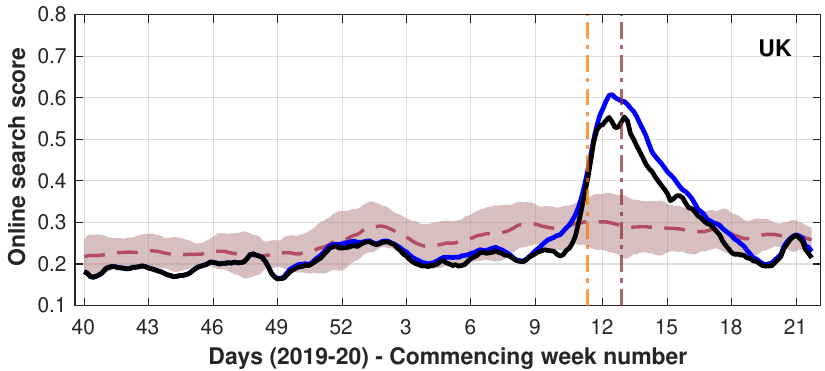}}{0in}{0in} \\
    
    & \def\stackalignment{l}
    \topinset{}{\includegraphics[height=1.2in, clip=true, trim=0.23in 0.35in 0 0]{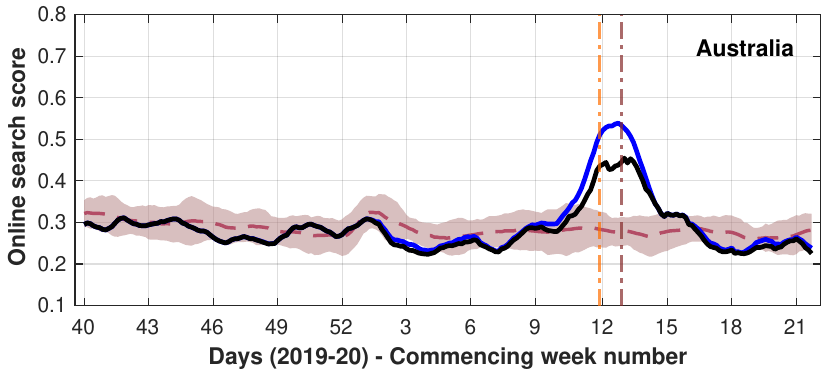}}{0in}{0in}
    \def\stackalignment{l}
    \topinset{}{\hspace{0.04in}\includegraphics[height=1.2in, clip=true, trim=0.45in 0.35in 0 0]{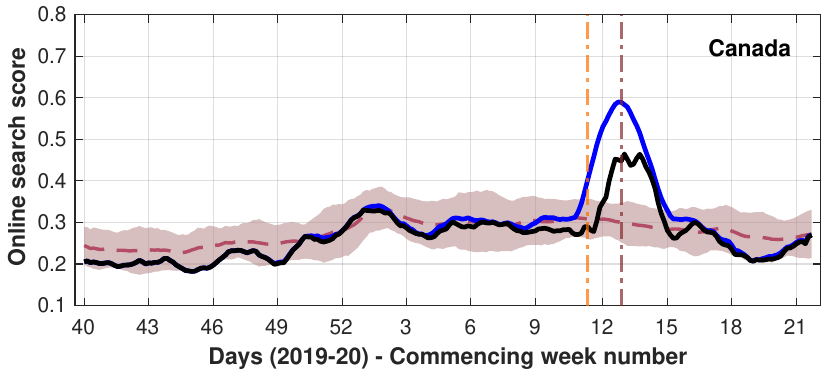}}{0in}{0in} \\
    
    & \def\stackalignment{l}
    \topinset{}{\includegraphics[height=1.2in, clip=true, trim=0.23in 0.35in 0 0]{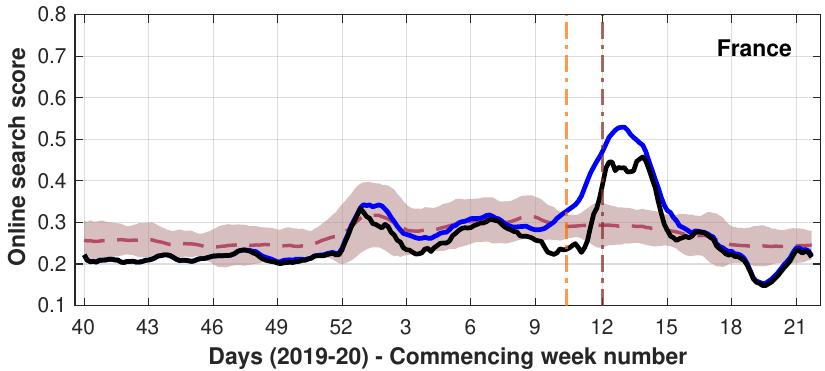}}{0in}{0in}
    \def\stackalignment{l}
    \topinset{}{\hspace{0.04in}\includegraphics[height=1.2in, clip=true, trim=0.45in 0.35in 0 0]{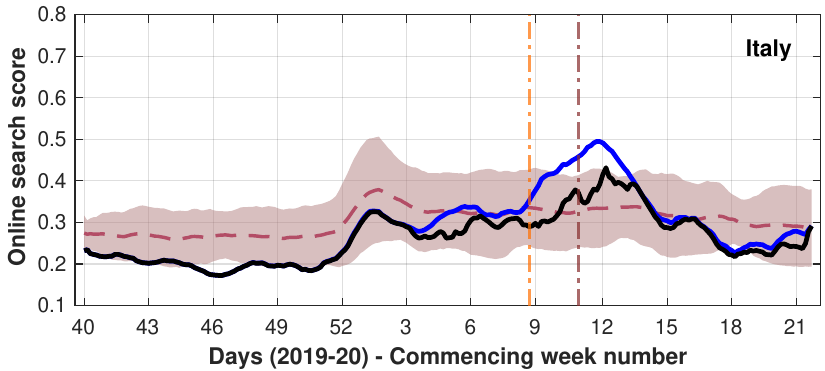}}{0in}{0in} \\
    
    & \def\stackalignment{l}
    \topinset{}{\includegraphics[height=1.285in, clip=true, trim=0.23in 0.20in 0 0]{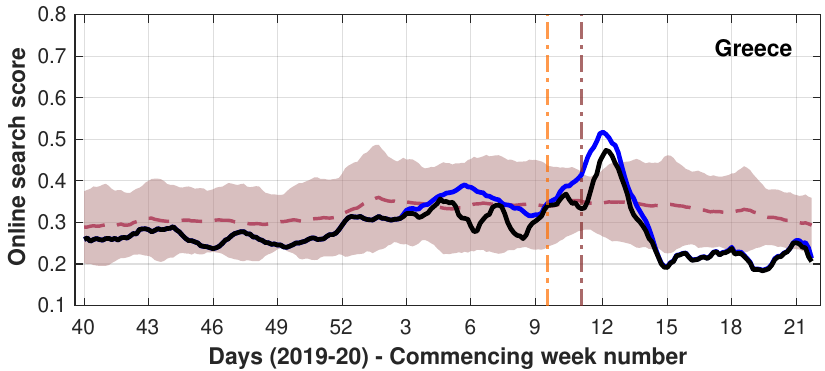}}{0in}{0in}
    \def\stackalignment{l}
    \topinset{}{\hspace{0.04in}\includegraphics[height=1.285in, clip=true, trim=0.45in 0.20in 0 0]{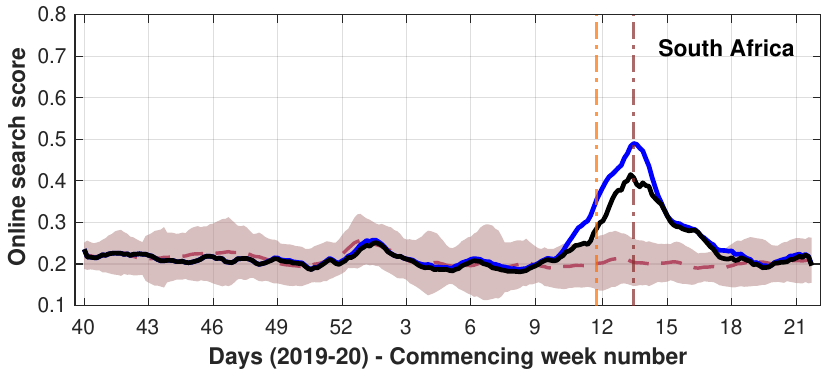}}{0in}{0in}\\
    
    \multicolumn{2}{c}{\sffamily\small\hspace{0.3in} Days (2019-20) -- Commencing week number}
    
    \end{tabular}
    
    \caption{Online search scores for COVID-19-related symptoms as identified by the FF100 survey, in addition to queries with coronavirus-related terms, for $8$ countries from September 30, 2019 to \analysisDate~(all inclusive). Query frequencies are uniformly weighted (blue line), and have news media effects minimised (black line). These scores are compared to an average $8$-year trend of the uniformly weighted model (dashed line) and its corresponding $95\%$ confidence intervals (shaded area). Application dates for physical distancing or lockdown measures are indicated with dash-dotted vertical lines; for countries that deployed different regional approaches, the first application of such measures is depicted. All time series are smoothed using a $7$-point moving average, centred around each day.}
    \label{fig:resultsNoWeights}
\end{figure}

\begin{figure}[ht]
    \centering
    \begin{tabular}{cc}
    \multirow{4}{*}[-2.2em]{\rotatebox[origin=c]{90}{\sffamily\small Standardised time series trend (z-score)}} & \def\stackalignment{l}
    \topinset{}{\includegraphics[height=1.2in, clip=true, trim=0.23in 0.35in 0 0]{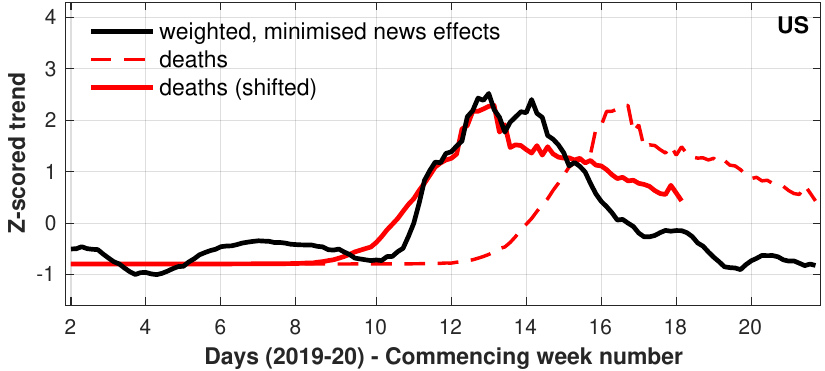}}{0in}{0in}
    \def\stackalignment{l}
    \topinset{}{\hspace{0.04in}\includegraphics[height=1.2in, clip=true, trim=0.38in 0.35in 0 0]{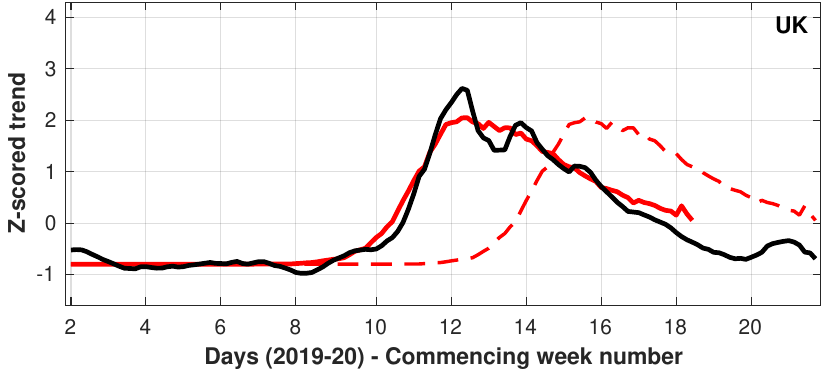}}{0in}{0in} \\
    
    & \def\stackalignment{l}
    \topinset{}{\includegraphics[height=1.2in, clip=true, trim=0.23in 0.35in 0 0]{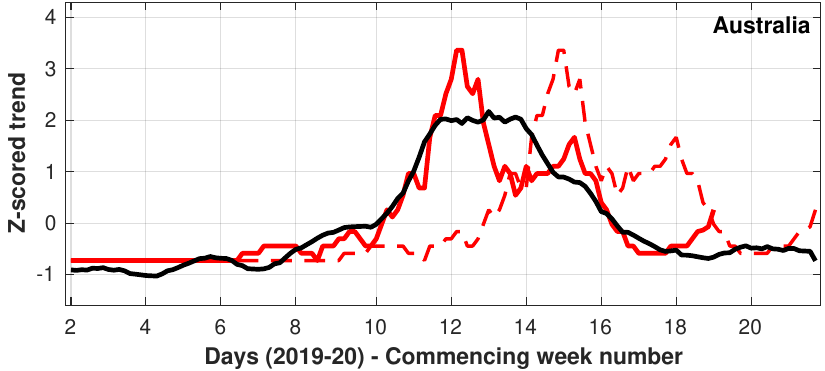}}{0in}{0in}
    \def\stackalignment{l}
    \topinset{}{\hspace{0.04in}\includegraphics[height=1.2in, clip=true, trim=0.38in 0.35in 0 0]{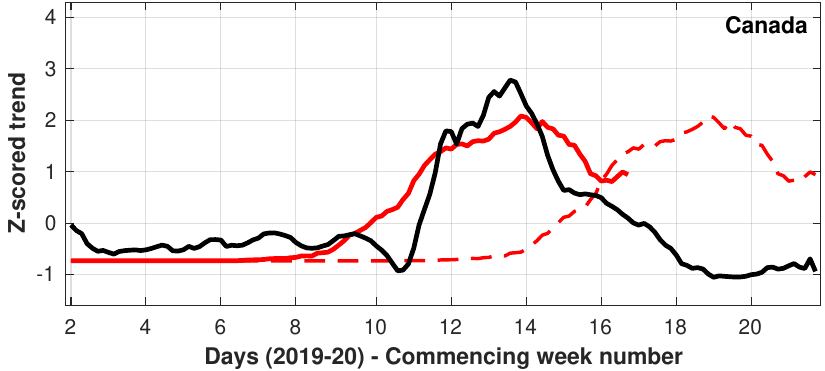}}{0in}{0in} \\
    
    & \def\stackalignment{l}
    \topinset{}{\includegraphics[height=1.2in, clip=true, trim=0.23in 0.35in 0 0]{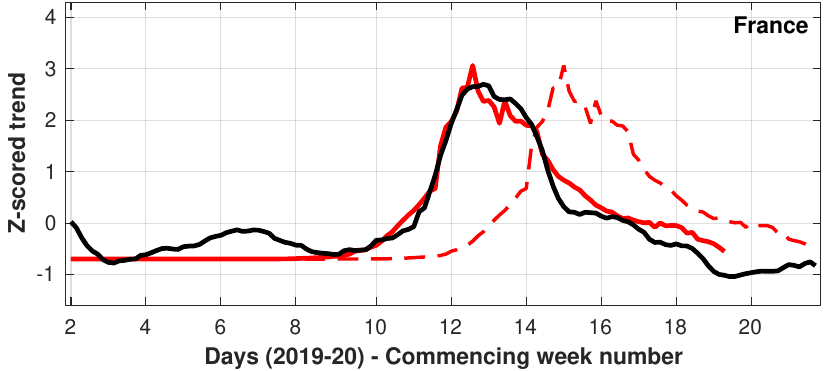}}{0in}{0in}
    \def\stackalignment{l}
    \topinset{}{\hspace{0.04in}\includegraphics[height=1.2in, clip=true, trim=0.38in 0.35in 0 0]{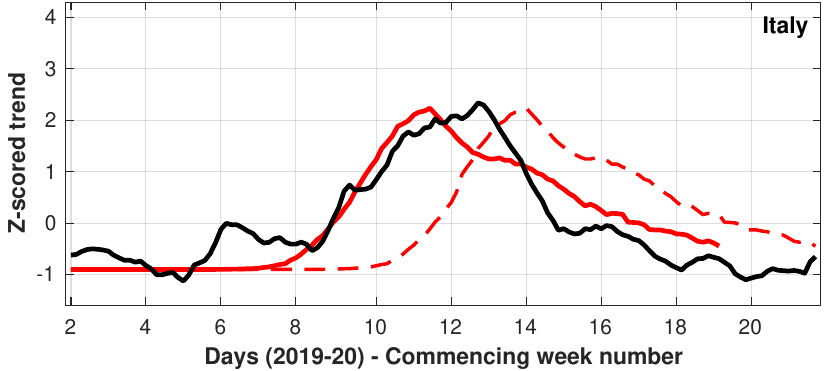}}{0in}{0in} \\
    
    & \def\stackalignment{l}
    \topinset{}{\includegraphics[height=1.285in, clip=true, trim=0.23in 0.20in 0 0]{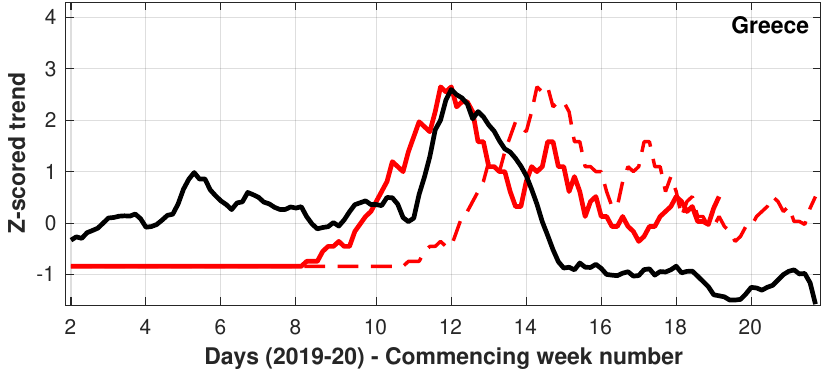}}{0in}{0in}
    \def\stackalignment{l}
    \topinset{}{\hspace{0.04in}\includegraphics[height=1.285in, clip=true, trim=0.38in 0.20in 0 0]{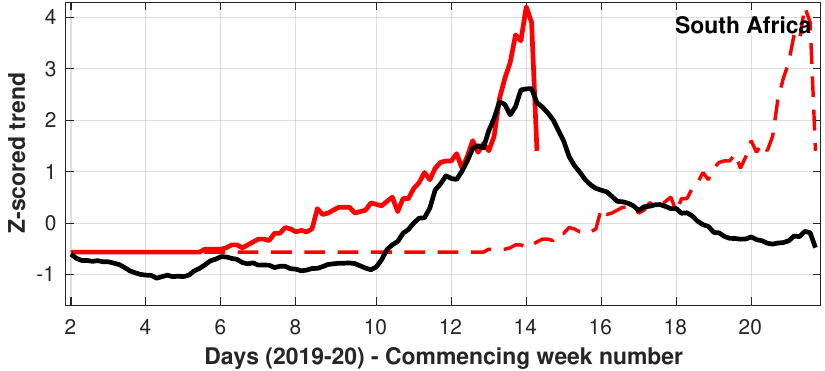}}{0in}{0in}\\
    
    \multicolumn{2}{c}{\sffamily\small\hspace{0.34in} Days (2019-20) -- Commencing week number}
    
    \end{tabular}
    
    \caption{Comparison between online search scores with minimised news media effects (black line) with deaths caused by COVID-19 (dashed red line), as well as deaths shifted back (red line) such that their correlation with the online search scores is maximised. The deaths time series are shifted back by a different number of days for each country: 25 days (US), 23 days (UK), 19 days (Australia), 35 days (Canada), 17 days (France), 18 days (Italy), 18 days (Greece), and 52 days (South Africa). The depicted time interval is from September 30, 2019 to \analysisDate~(all inclusive). All time series are smoothed using a $7$-point moving average, centred around each day.}
    \label{fig:uns_vs_d}
\end{figure}

Fig.~\ref{fig:searchVol} compares the frequency of search queries that include the words ``the'', ``weather'' or ``coronavirus'' in the United States (US) and the United Kingdom (UK) to illustrate the unprecedented characteristics in the volume of online searches during the COVID-19 pandemic. Fig.~\ref{fig:results} depicts a similar result to Fig.~1, with the difference that the scores are solely based on the first few hundred (FF100) symptom categories~\cite{Boddington2020} without including generic queries about coronavirus. Fig.~\ref{fig:resultsNoWeights} is also similar to Fig.~1, but this time the weighting of queries is uniform (equal weights for all symptoms). Fig.~\ref{fig:uns_vs_d} shows a comparison between the unsupervised model with minimised news media effects from Fig.~1 and the corresponding deaths caused by COVID-19; it also shifts back the deaths time series to maximise the correlation between the two signals. Fig.~\ref{fig:results_tl_sup} Fig.~\ref{fig:results_tl} shows progressive versions of Fig.~3, i.e. all intermediate model outputs (training was conducted on a daily basis). Fig.~\ref{fig:feature_analysis_B} shows additional results from the correlation and regression analysis. Figs.~\ref{fig:results_forecasting_deaths},~\ref{fig:results_forecasting_cc}, and Table~\ref{tab:forecastingAC_sup}, show the performance results and corresponding plots for the forecasting tasks of deaths and confirmed COVID-19 cases. Finally, Fig.~\ref{fig:news_ratio} shows the average daily news articles proportion about COVID-19 across all countries included in our analysis.


\newpage

\begin{table}[ht]
\footnotesize
\renewcommand*{\arraystretch}{1.4}
\setlength{\belowrulesep}{0pt}
\setlength{\aboverulesep}{0pt}
\setlength{\tabcolsep}{5pt}
\centering
\begin{tabular}{l rrr rrr}
\toprule
\multirow{2}{*}{\textbf{Country}} & \multicolumn{3}{c}{\bf 7 days ahead} & \multicolumn{3}{c}{\bf 14 days ahead} \\
&\multicolumn{1}{c}{\bf AR-F} &\multicolumn{1}{c}{\bf SAR-F} &\multicolumn{1}{c}{\bf PER-F} &\multicolumn{1}{c}{\bf AR-F} &\multicolumn{1}{c}{\bf SAR-F} &\multicolumn{1}{c}{\bf PER-F} \\
\midrule
\bf UK & 1672.20 (1235.59) & 1046.45 (1035.15) & 846.94 (830.41) & 2326.82 (1464.47) & 1525.97 (1381.28) & 1256.80 (1273.08) \\
\bf US & 6090.02 (4159.88) & 5147.06 (4188.73) & 5053.74 (4775.49) & 11037.60 (8624.17) & 6703.70 (7295.66) & 5665.03 (5423.37) \\
\bf Australia & 123.59 (90.33) & 42.62 (42.89) & 11.80 (9.78) & 217.74 (92.53) & 192.63 (114.35) & 24.86 (27.30) \\
\bf Canada & 392.17 (247.74) & 426.98 (358.46) & 273.06 (329.19) & 717.30 (444.85) & 414.10 (341.87) & 414.83 (320.46) \\
\bf Greece & 30.70 (17.05) & 28.03 (18.32) & 22.37 (32.19) & 40.80 (18.46) & 25.97 (19.02) & 23.37 (27.27) \\
\bf Italy & 1297.69 (508.56) & 681.91 (417.17) & 604.20 (311.72) & 2782.43 (1189.31) & 1603.71 (845.22) & 1149.54 (506.22) \\
\bf France & 1058.79 (717.59) & 824.56 (552.89) & 675.14 (851.49) & 2050.87 (953.91) & 1518.12 (586.42) & 1210.69 (1022.56) \\
\bf South Africa & 374.96 (241.57) & 297.92 (211.29) & 186.34 (141.94) & 448.41 (307.28) & 407.52 (253.51) & 299.26 (196.18) \\
\midrule
\bf Norm. mean & 0.231 (0.161) & 0.164 (0.143) & 0.119 (0.143) & 0.387 (0.235) & 0.271 (0.204) & 0.183 (0.172) \\
\bottomrule
\end{tabular}
\caption{\label{tab:forecastingAC_sup} Average mean absolute error and standard deviation (in parentheses) of forecasting models ($7$ and $14$ days ahead) for daily confirmed COVID-19 cases in 8 countries. The last row contains min-max normalised averages across countries, methods, and forecasting tasks to account for the different ranges in different countries. \textbf{AR-F}: autoregressive forecasting using past confirmed cases; \textbf{SAR-F}: combined online search and autoregressive forecasting; \textbf{PER-F}: persistence model.}
\end{table}

\vspace{1in}

\newpage

\begin{figure}[ht]
    \centering
    \setlength{\tabcolsep}{0.1em}
    \begin{tabular}{ccc}
    
    \def\stackalignment{l}
    \topinset{}{\includegraphics[height=1in, clip=true, trim=0.25in 0.5in 0 0]{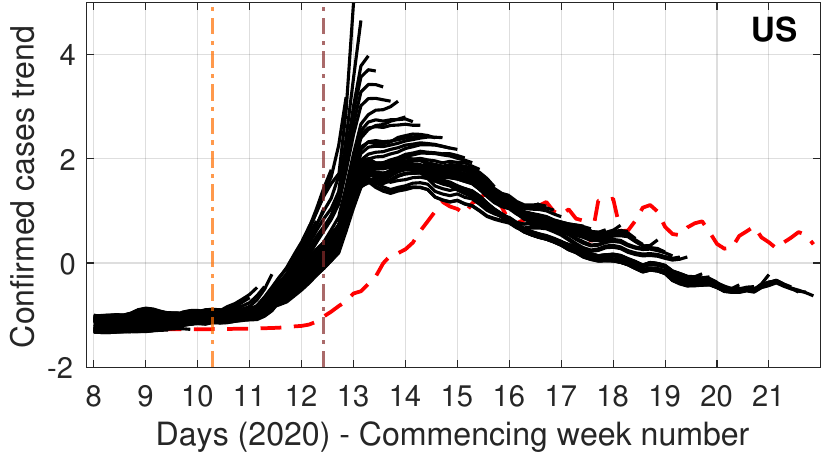}}{0in}{0in} &
    \def\stackalignment{l}
    \topinset{}{\includegraphics[height=1in, clip=true, trim=0.52in 0.5in 0 0]{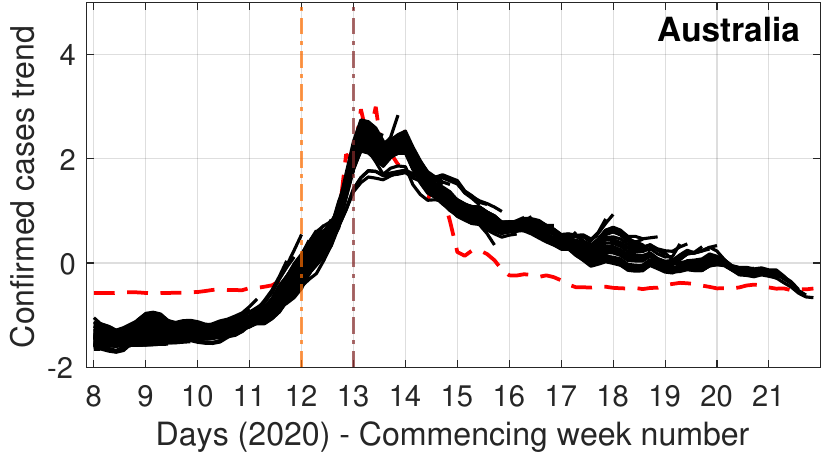}}{0in}{0in} &
    \def\stackalignment{l}
    \topinset{}{\includegraphics[height=1in, clip=true, trim=0.52in 0.5in 0 0]{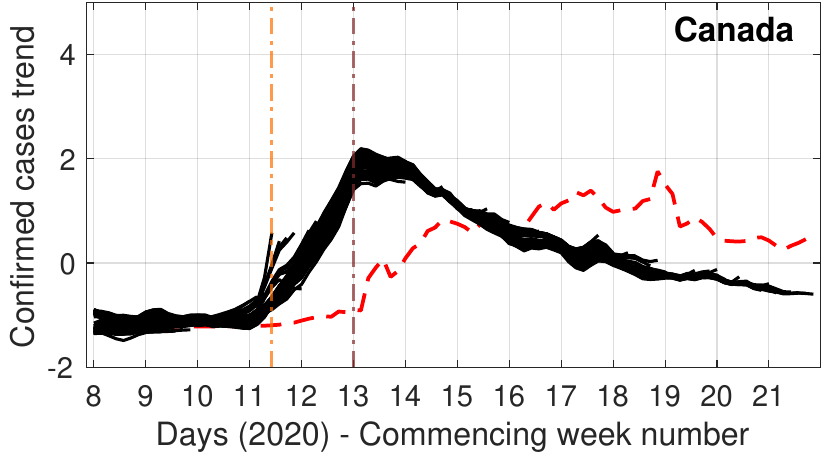}}{0in}{0in} \\
    
    \def\stackalignment{l}
    \topinset{}{\includegraphics[height=1.095in, clip=true, trim=0.25in 0.25in 0 0]{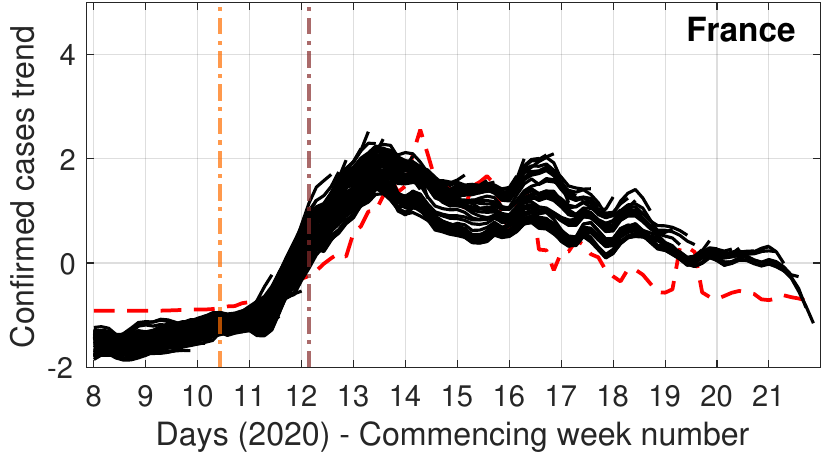}}{0in}{0in} &
    \def\stackalignment{l}
    \topinset{}{\includegraphics[height=1.095in, clip=true, trim=0.52in 0.25in 0 0]{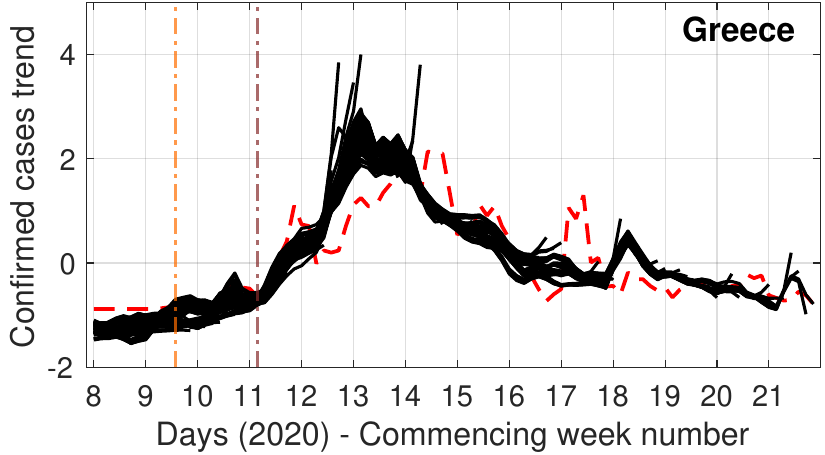}}{0in}{0in} &
    \def\stackalignment{l}
    \topinset{}{\includegraphics[height=1.095in, clip=true, trim=0.52in 0.25in 0 0]{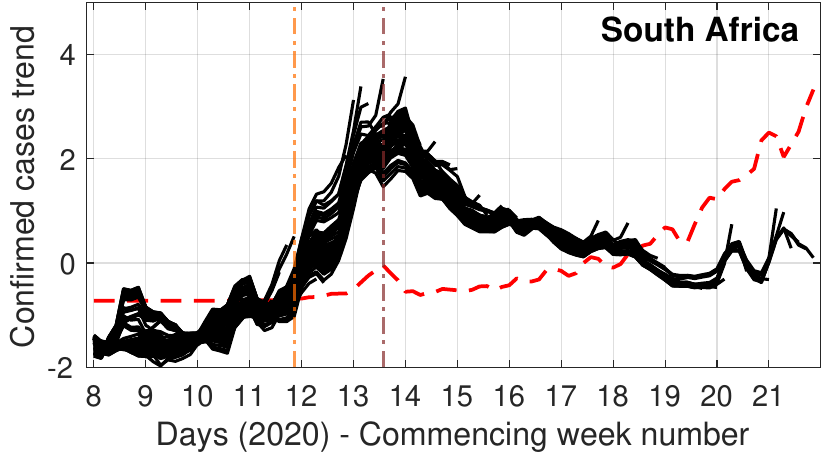}}{0in}{0in} \\
    
    \multicolumn{3}{c}{\def\stackalignment{l}
    \topinset{}{\includegraphics[height=2.1in, clip=true, trim=0 0 0 0]{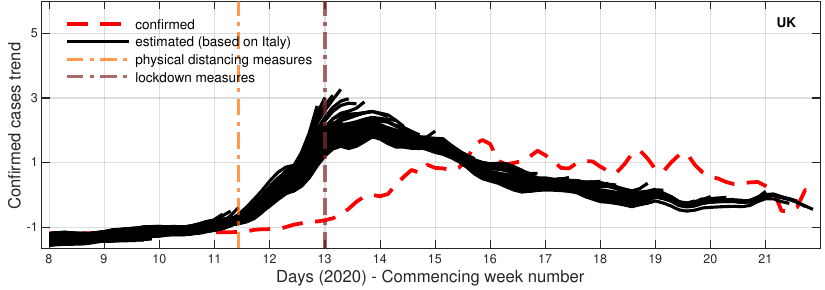}}{0in}{0in}}
    
    \end{tabular}
    
    \caption{Transfer learning models for $7$ countries and their temporal progression using Italy as the source country. The figures show an ongoing (updated on a daily basis) estimated trend for confirmed COVID-19 cases compared to the reported one. The solid line represents the mean estimate from an ensemble of models. Application dates for physical distancing or lockdown measures are indicated with dash-dotted vertical lines; for countries that deployed different regional approaches, the first application of such measures is depicted. Time series are standardised and smoothed using a 3-point moving average, centred around each day.}
    \label{fig:results_tl_sup}
\end{figure}

\begin{figure}[ht]
    \centering
    \def\stackalignment{l}
    \topinset{\sffamily\textbf{A}}{\includegraphics[height=3.4in, clip=true, trim=0 0 0 0]{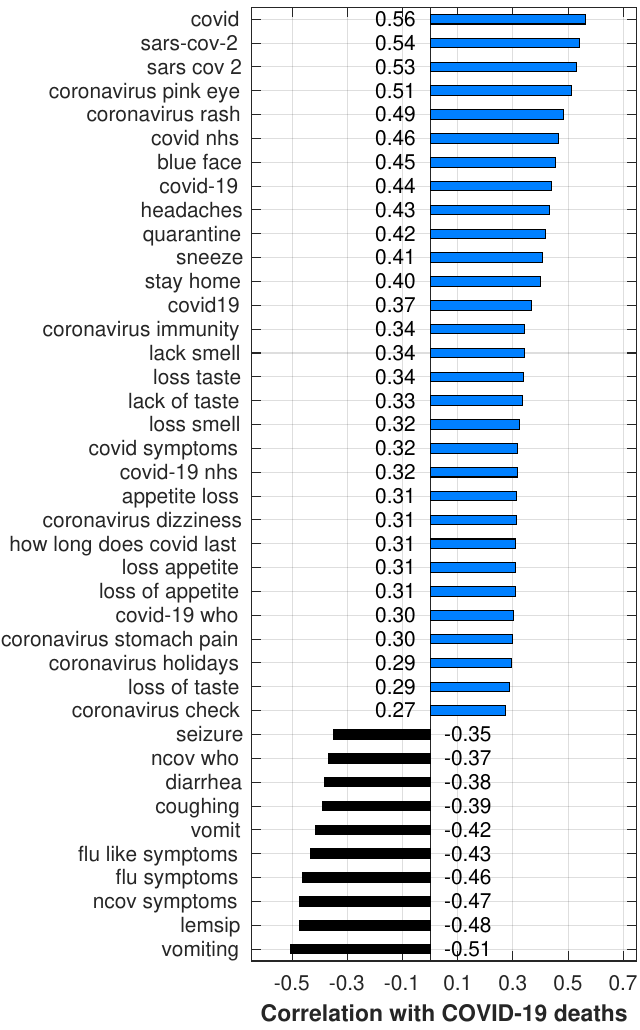}}{0.01in}{0.1in}
    \hspace{0.1in}
    \def\stackalignment{l}
    \topinset{\sffamily\textbf{B}}{\includegraphics[height=3.4in, clip=true, trim=0 0 0 0]{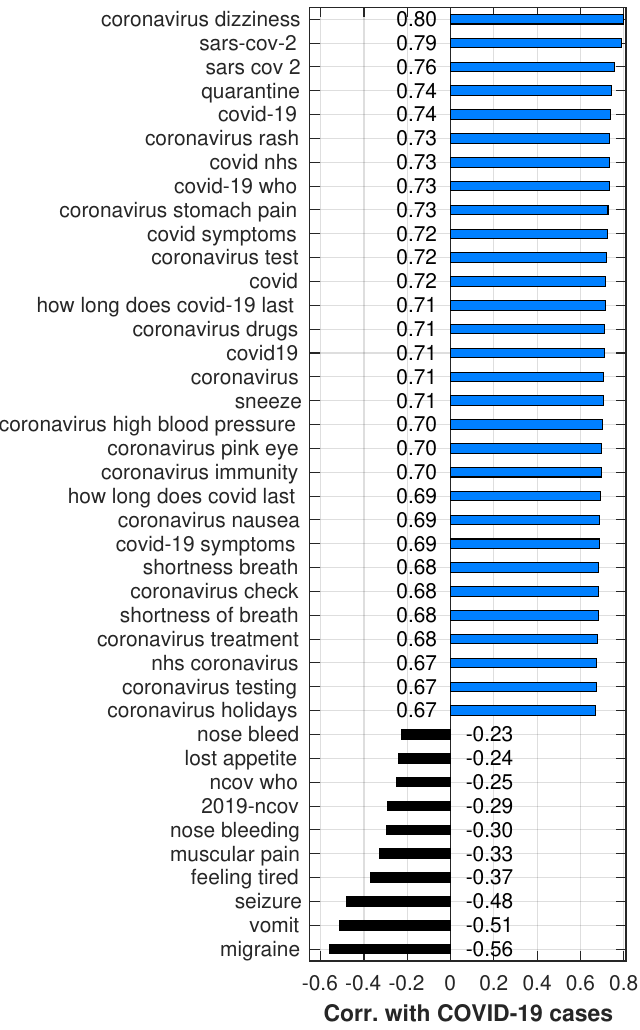}}{0.01in}{0.1in}
    \hspace{0.1in}
    \def\stackalignment{l}
    \topinset{\sffamily\textbf{C}}{\includegraphics[height=3.4in, clip=true, trim=0 0 0 0]{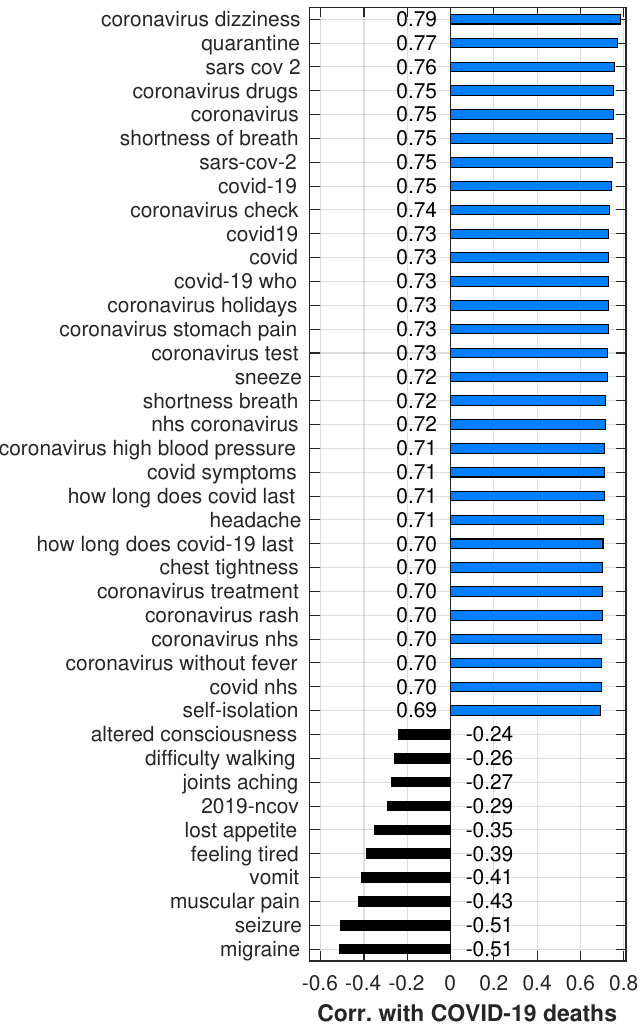}}{0.01in}{0.1in}
    \\
    \vspace{0.2in}
    
    \def\stackalignment{l}
    \topinset{\sffamily\textbf{D}}{\includegraphics[height=3.4in, clip=true, trim=0 0 0 0]{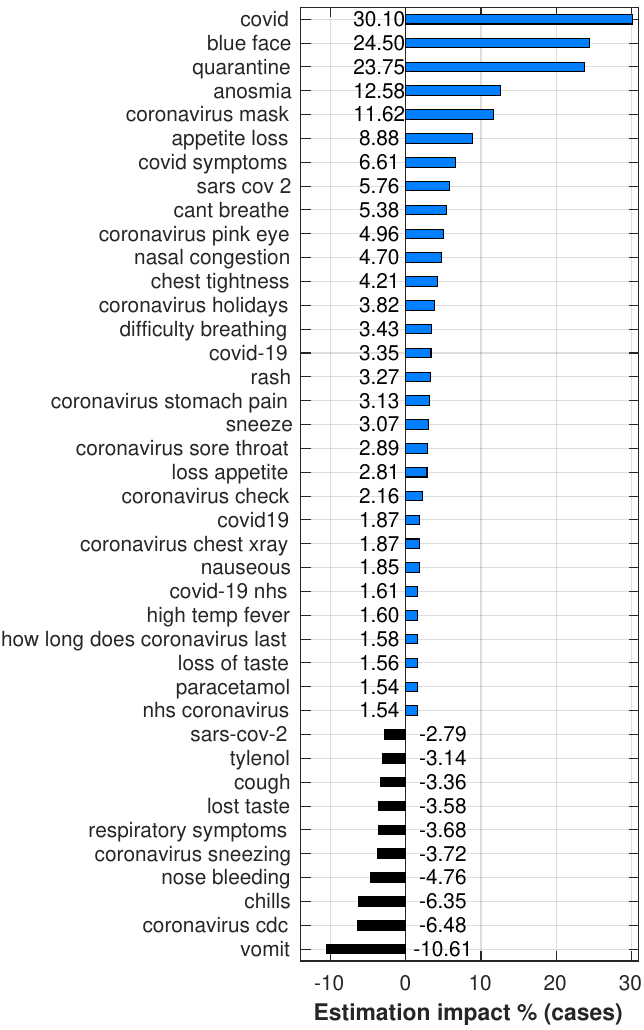}}{0.01in}{0.1in}
    \hspace{0.1in}
    \def\stackalignment{l}
    \topinset{\sffamily\textbf{E}}{\includegraphics[height=3.4in, clip=true, trim=0 0 0 0]{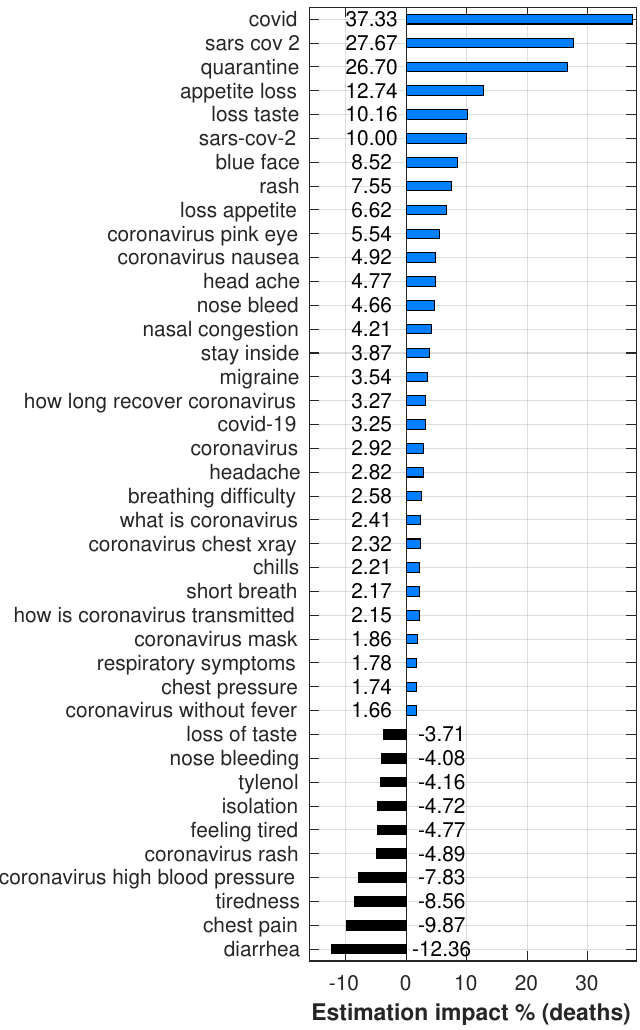}}{0.01in}{0.1in}
    
    \caption{Correlation and regression analysis of search query frequencies against confirmed COVID-19 cases or deaths in four English speaking countries (US, UK, Australia, and Canada). (\textbf{A}) Top-30 positively and top-10 negatively correlated search queries with deaths caused by COVID-19; (\textbf{B}) Top-30 positively and top-10 negatively correlated search queries with confirmed COVID-19 cases after bringing their time series forward by 19 days (average maximum correlation); (\textbf{C}) Top-30 positively and top-10 negatively correlated search queries with deaths caused by COVID-19 after bringing their time series forward by 25 days (average maximum correlation); (\textbf{D}) Top-30 positively and top-10 negatively impactful queries in estimating COVID-19 confirmed cases while exploring the entire regularisation path; (\textbf{E}) Top-30 positively and top-10 negatively impactful queries in estimating deaths caused by COVID-19 while exploring the entire regularisation path.}
    \label{fig:feature_analysis_B}
\end{figure}

\begin{figure}[ht]
    \centering
    \begin{tabular}{cc}
    \multirow{4}{*}[2.95em]{\rotatebox[origin=c]{90}{\sffamily\small Deaths (normalised)}} & \def\stackalignment{l}
    \topinset{}{\includegraphics[height=0.88in, clip=true, trim=0.25in 0.46in 0 0]{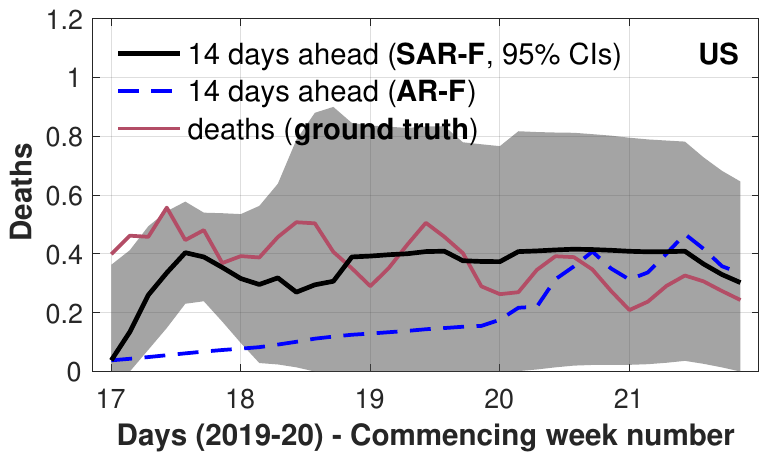}}{0in}{0in}
    \def\stackalignment{l}
    \topinset{}{\includegraphics[height=0.88in, clip=true, trim=0.6in 0.46in 0 0]{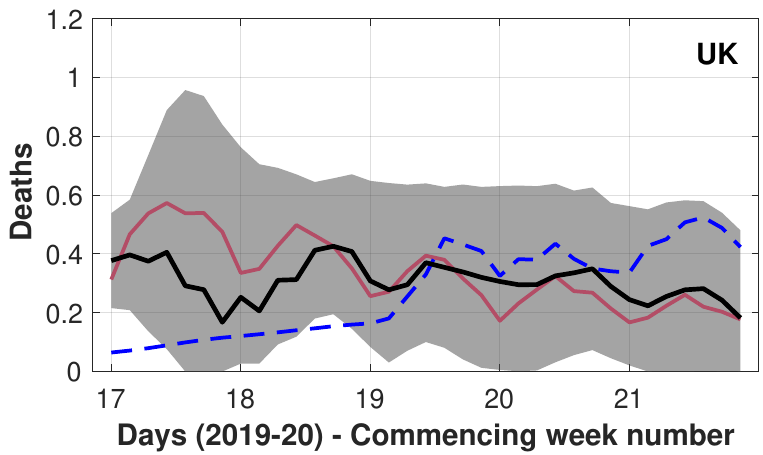}}{0in}{0in} 
    \def\stackalignment{l}
    \topinset{}{\includegraphics[height=0.88in, clip=true, trim=0.6in 0.46in 0 0]{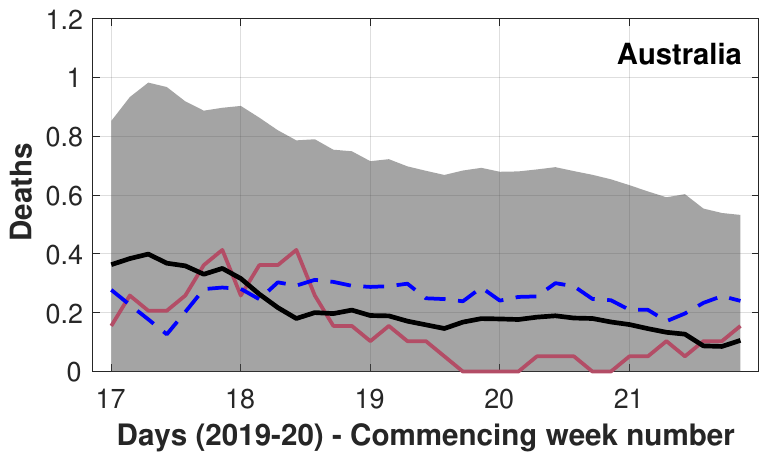}}{0in}{0in}
    \def\stackalignment{l}
    \topinset{}{\includegraphics[height=0.88in, clip=true, trim=0.6in 0.46in 0 0]{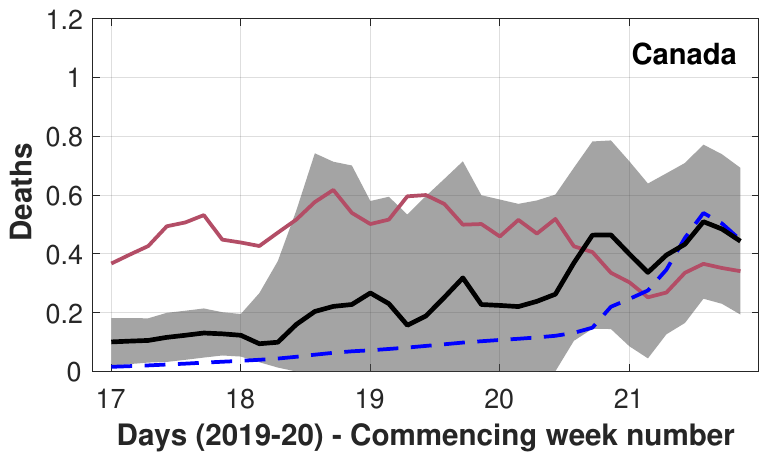}}{0in}{0in}
    \\
    
    & 
    \def\stackalignment{l}
    \topinset{}{\includegraphics[height=0.934in, clip=true, trim=0.25in 0.30in 0 0]{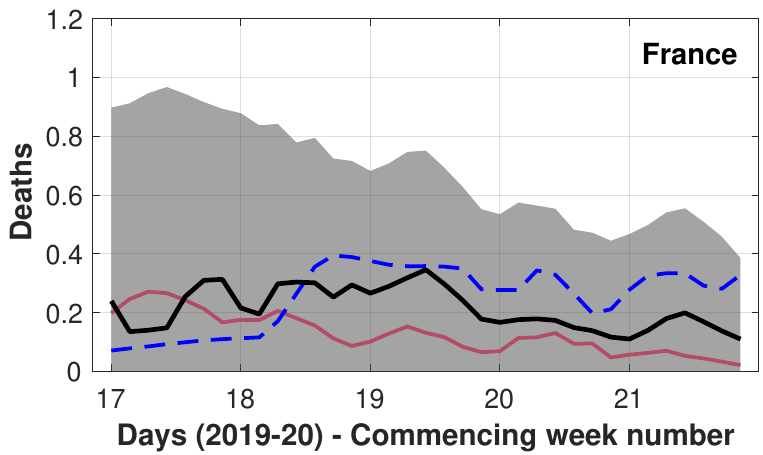}}{0in}{0in}
    \def\stackalignment{l}
    \topinset{}{\includegraphics[height=0.934in, clip=true, trim=0.6in 0.30in 0 0]{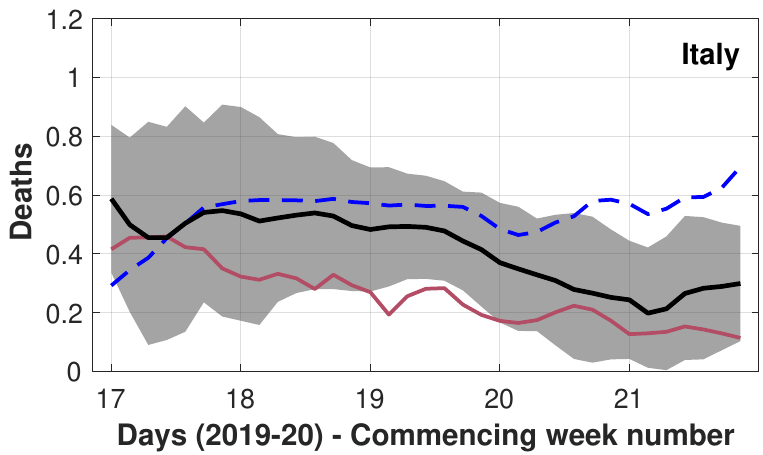}}{0in}{0in}
    \def\stackalignment{l}
    \topinset{}{\includegraphics[height=0.934in, clip=true, trim=0.6in 0.30in 0 0]{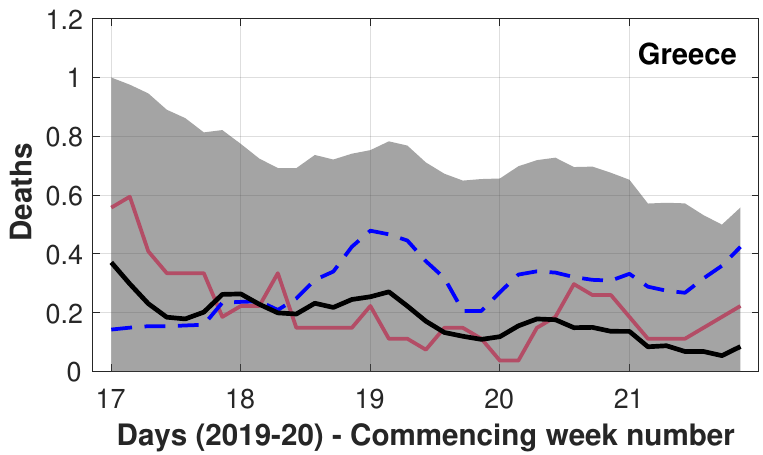}}{0in}{0in}
    \def\stackalignment{l}
    \topinset{}{\includegraphics[height=0.934in, clip=true, trim=0.6in 0.30in 0 0]{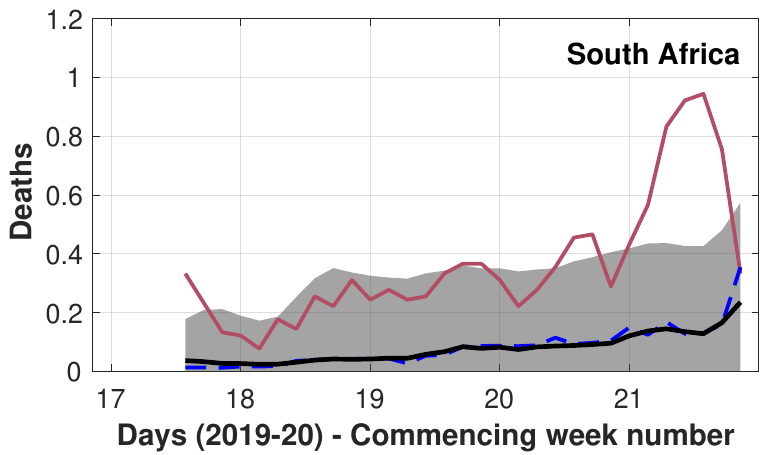}}{0in}{0in}\\
    
    \multicolumn{2}{c}{\sffamily\small\hspace{0.35in} Days (2020) -- Commencing week number}
    
    \end{tabular}

    \caption{$14$-days ahead daily forecasting estimates of deaths caused by COVID-19 for $8$ countries starting from April 20, 2020 or from the date by which a cumulative number of $10$ deaths have been reported (differs for South Africa). Deaths are depicted using a red line. The dashed blue line shows deaths forecasts from a strictly autoregressive model (\textbf{AR-F}). The black line shows deaths forecasts from a model that incorporates online search information (\textbf{SAR-F}). The shaded area denotes the corresponding 95\% confidence intervals for the latter estimates. For a better visualisation, all values are normalised using min-max, and are smoothed using a 3-point moving average, centred around each day.}
    \label{fig:results_forecasting_deaths}
\end{figure}

\begin{figure}[ht]
    \centering
    \begin{tabular}{cc}
    \multirow{4}{*}[5em]{\rotatebox[origin=c]{90}{\sffamily\small Confirmed cases (normalised)}} & \def\stackalignment{l}
    \topinset{}{\includegraphics[height=0.88in, clip=true, trim=0.25in 0.46in 0 0]{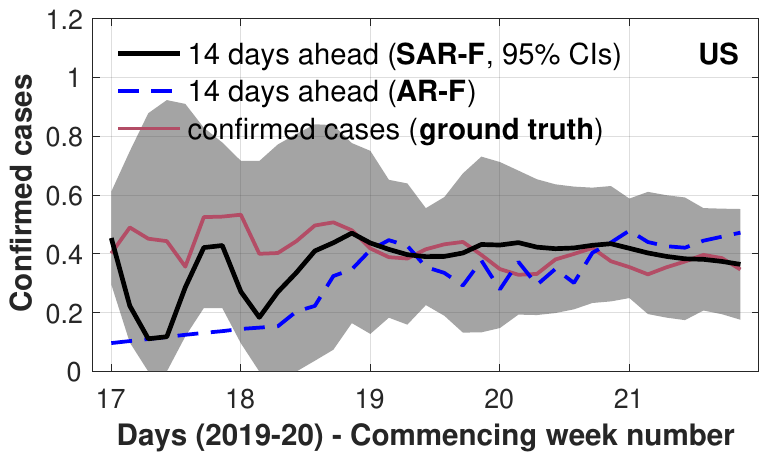}}{0in}{0in}
    \def\stackalignment{l}
    \topinset{}{\includegraphics[height=0.88in, clip=true, trim=0.6in 0.46in 0 0]{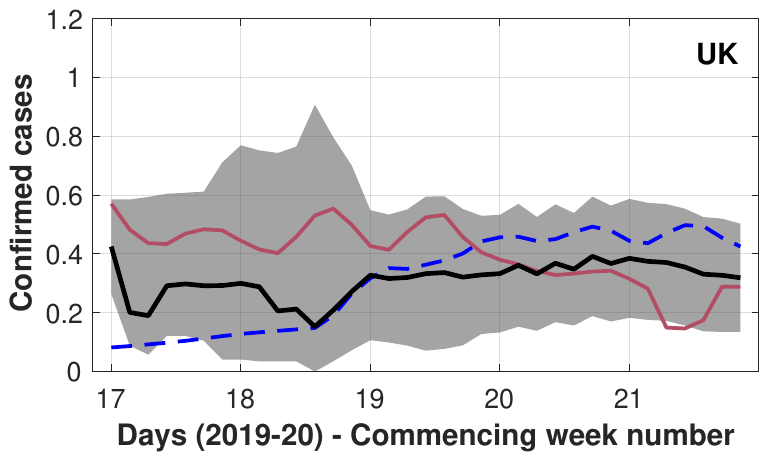}}{0in}{0in} 
    \def\stackalignment{l}
    \topinset{}{\includegraphics[height=0.88in, clip=true, trim=0.6in 0.46in 0 0]{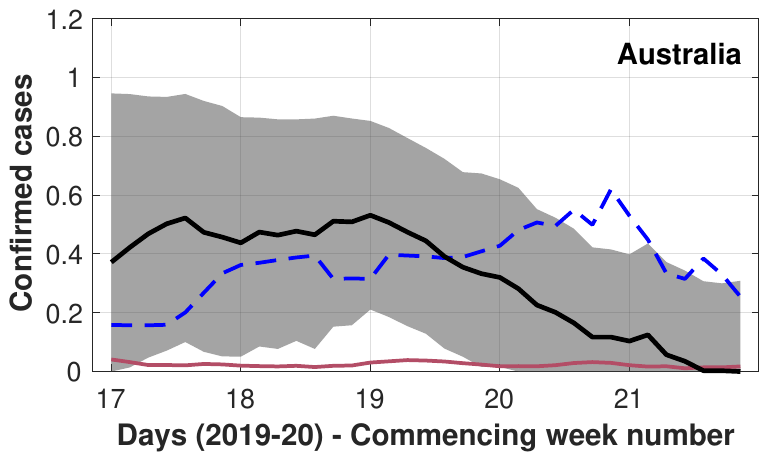}}{0in}{0in}
    \def\stackalignment{l}
    \topinset{}{\includegraphics[height=0.88in, clip=true, trim=0.6in 0.46in 0 0]{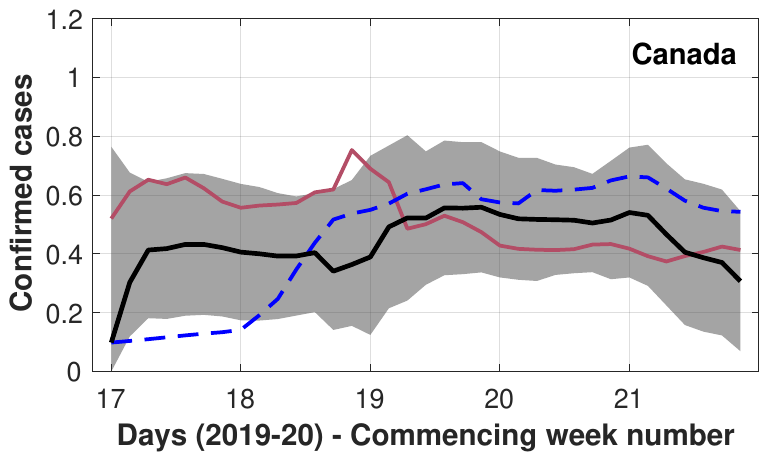}}{0in}{0in}
    \\
    
    & 
    \def\stackalignment{l}
    \topinset{}{\includegraphics[height=0.934in, clip=true, trim=0.25in 0.30in 0 0]{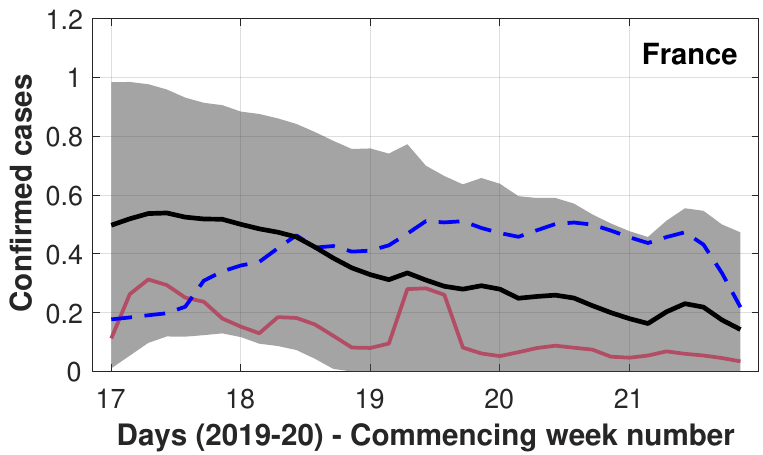}}{0in}{0in}
    \def\stackalignment{l}
    \topinset{}{\includegraphics[height=0.934in, clip=true, trim=0.6in 0.30in 0 0]{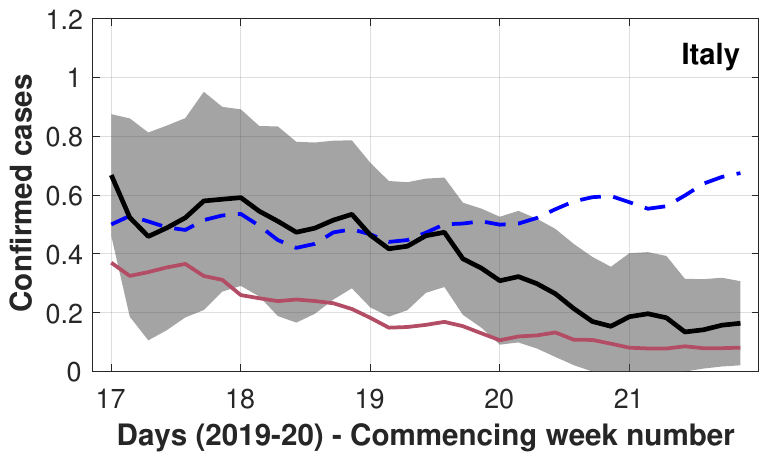}}{0in}{0in}
    \def\stackalignment{l}
    \topinset{}{\includegraphics[height=0.934in, clip=true, trim=0.6in 0.30in 0 0]{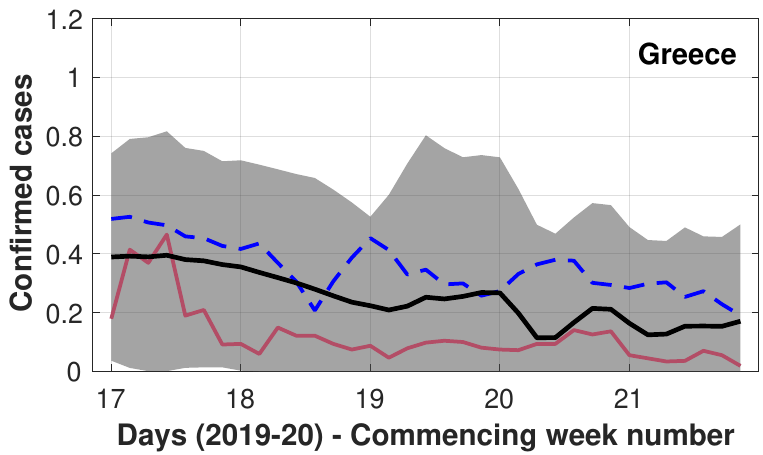}}{0in}{0in}
    \def\stackalignment{l}
    \topinset{}{\includegraphics[height=0.934in, clip=true, trim=0.6in 0.30in 0 0]{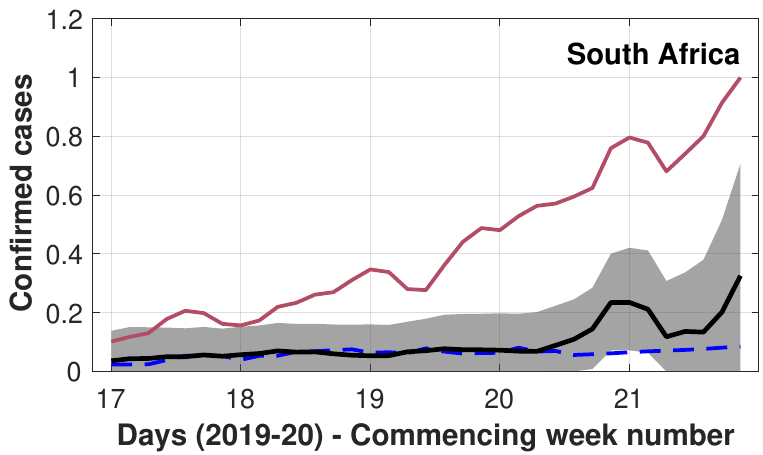}}{0in}{0in}\\
    
    \multicolumn{2}{c}{\sffamily\small\hspace{0.35in} Days (2020) -- Commencing week number}
    
    \end{tabular}

    \caption{$14$-days ahead daily forecasting estimates of confirmed COVID-19 cases for $8$ countries starting from April 20, 2020 or from the date by which a cumulative number of $250$ cases have been reported (different date for each country). Deaths are depicted using a red line. The dashed blue line shows deaths forecasts from a strictly autoregressive model (\textbf{AR-F}). The black line shows deaths forecasts from a model that incorporates online search information (\textbf{SAR-F}). The shaded area denotes the corresponding 95\% confidence intervals for the latter estimates. For a better visualisation, all values are normalised using min-max, and are smoothed using a 3-point moving average, centred around each day.}
    \label{fig:results_forecasting_cc}
\end{figure}

\begin{figure}[ht]
    \centering
    \includegraphics[width=0.95\textwidth]{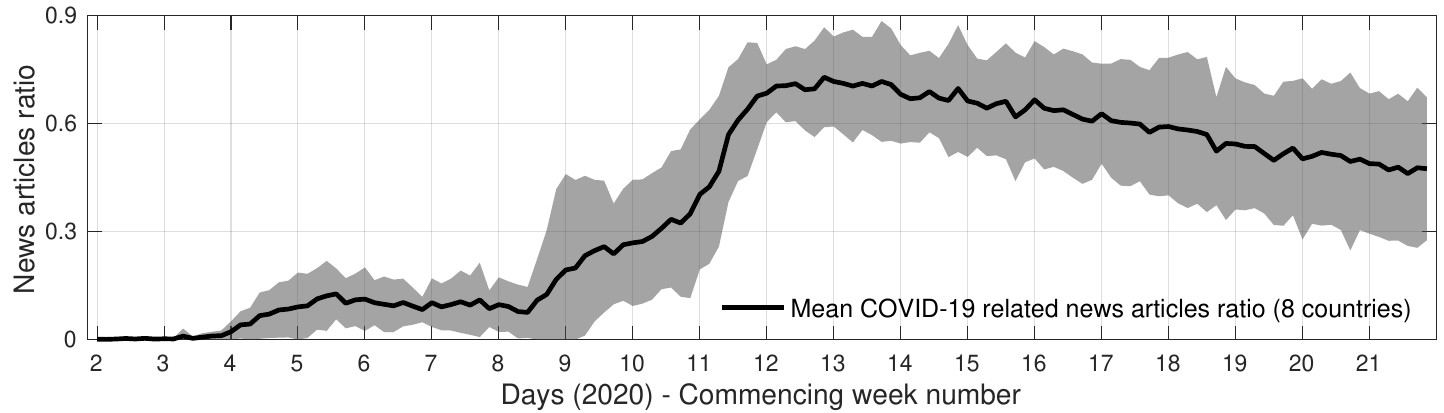}
    \caption{Average daily news articles proportion about COVID-19 across all countries in our analysis and corresponding confidence intervals (two standard deviations above and below the mean).}
    \label{fig:news_ratio}
\end{figure}

\end{document}